\def\bar#1{\overline{#1}}

\def\Hat#1{\rlap{\kern.10em$\widehat{\phantom G}$}#1}
\def\HAt#1{\rlap{\kern.05em$\widehat{\phantom G}$}#1}

\def\cap#1{\rlap{\kern.1em$\widehat{\phantom{G\vrule height.8em}}$}#1{}}
\def\Cap#1{\rlap{\kern.05em$\widehat{\phantom{G\vrule height.8em}}$}#1{}}

\documentstyle[12pt]{article}
\normalsize

\let\oldtheequation=\theequation
\def\doteqs#1{\setcounter{equation}{0}
            \def\theequation{{#1}.\oldtheequation}}

\newcounter{sxn}
\def\sx#1{\addtocounter{sxn}{1} \bigskip\medskip \goodbreak \noindent{\large\bf
\centerline{\thesxn.~~#1}} \nobreak \medskip}
\def\sxn#1{\sx{#1} \doteqs{\thesxn}}

\newcounter{axn}

\def\br{\begin{thebibliography}{abc}}
\def\er{\end{thebibliography}}

\date{}

\begin{document}
\bibliographystyle{unsrt}
\footskip 1.0cm
\thispagestyle{empty}
\begin{flushright}
UAHEP 925\\
SU-4240-508\\
INFN-NA-IV-92/12\\
May 1992\\
\end{flushright}
\begin{center}{\LARGE EDGE CURRENTS AND\\
         VERTEX OPERATORS\\
         FOR CHERN-SIMONS GRAVITY\\}
\vspace*{10mm}
{\large   G. Bimonte $^{1,2}$
          K. S. Gupta,$^{1}$
          A. Stern $^{2,3}$ \\ }
\newcommand{\bc}{\begin{center}}
\newcommand{\ec}{\end{center}}
\vspace*{10mm}
 1){\it Department of Physics, Syracuse University,\\
Syracuse, NY 13244-1130, USA}.\\
\vspace*{5mm}
 2){\it Dipartimento di Scienze Fisiche dell' Universit\`a di Napoli,\\
    Mostra d'Oltremare pad. 19, 80125 Napoli, Italy}.\\
\vspace*{5mm}
 3){\it Department of Physics, University of Alabama, \\
Tuscaloosa, Al 35487, USA.}\ec

\newpage
\vspace*{7mm}
\normalsize
\centerline{\bf ABSTRACT}
\vspace*{5mm}

We apply elementary canonical methods for the quantization of 2+1
dimensional gravity, where the dynamics is given by E. Witten's
$ISO(2,1)$ Chern-Simons action.  As in a previous work,
our approach does not involve choice of gauge or clever
manipulations of functional integrals.
Instead, we just require the
Gauss law constraint for gravity to be first class and also to be
everywhere differentiable.
When the spatial slice is a disc, the gravitational fields can either
be unconstrained or constrained at the boundary of the disc.
The unconstrained fields correspond to edge currents which
carry a representation of the $ISO(2,1)$ Kac-Moody algebra.
Unitary representations for such an algebra have
been found using the method of induced representations.
In the case of constrained fields, we can classify all possible
boundary conditions.  For several different boundary conditions, the
field content of the theory reduces precisely to
that of 1+1 dimensional gravity theories.  We extend the above
formalism to include sources.  The sources take into account self-
interactions.  This is done by
punching holes in the disc, and erecting an $ISO(2,1)$ Kac-Moody algebra
on the boundary of each hole.  If the hole is originally sourceless,
a source can be created via the action of a vertex operator $V$.  We give
an explicit expression for $V$.  We shall show that when acting
on the vacuum state, it creates particles with a discrete mass spectrum.
The lowest mass particle induces a cylindrical space-time geometry,
while higher mass particles give an $n$-fold covering of the
cylinder.  The vertex operator therefore creates cylindrical space-time
geometries from the vacuum.
\newpage

\newcommand{\be}{\begin{equation}}
\newcommand{\ee}{\end{equation}}

\baselineskip=24pt
\setcounter{page}{1}
\sxn{INTRODUCTION}

Even though there are no gravitons in three dimensional gravity,
the theory can have a rich structure when the topology of space-time
is nontrivial.  This is especially evident in the Chern-Simons
formulation of the theory, proposed by E. Witten \cite{wit}.
In this formulation, the action is written in terms of
connection one forms $e^a$ and $\omega^a$ for the $ISO(2,1)$
group.  $ISO(2,1)$ denotes the Poincare group in three dimensions.
The components of $e^a$ are the dreibein fields from which one can
construct the space-time metric, while $\omega^a$ are the spin
connections from which one constructs the $SO(2,1)$ curvature two form.

In Chern-Simons theory,
when the space-time manifold is a disc $D \times {\bf R}^{1}$
(${\bf R}^{1}$ accounting for time) it is possible to quantize the
system in a manner which eliminates degrees of freedom from
the interior of the disc.  This also was proposed by E. Witten
\cite{wit2} and examined further by a number of authors [3-10].
(Elementary canonical methods were used in the approach of
A. P. Balachandran and us \cite{cs1,cs2},
and they did not require clever
manipulations of functional integrals or the imposition of
gauge constraints.  We believe that the methods of refs.
\cite{cs1,cs2} have the virtue of simplicity and we shall apply them
here to the case of the Chern-Simons description of $2+1$ gravity.)
After eliminating the degrees of freedom from the
interior of the disc, one is left with states associated
with the boundary $\partial D$, the so-called ``edge states".
In $U(1)$ Chern-Simons theory, these edge states play an important
role in the description of the quantum Hall effect \cite{zhang,blok}.
There, they
carry the unitary representations of $U(1)$ Kac-Moody algebras.
For the case of the Chern-Simons description of $2+1$ gravity,
the analogous of the edge states carry unitary representations of
the $ISO(2,1)$ Kac-Moody algebra.
$ISO(2,1)$ is a noncompact semidirect product group.
Highest weight representations \cite{go}, which are standardly employed
for the purpose of finding unitary representations of Kac-Moody
algebras associated with compact groups,
are plagued with difficulties in this case \cite{sss}.
Nevertheless, unitary representations have been found for the $ISO(2,1)$
Kac-Moody algebra \cite{sss}.  For them one applies
the method of induced representations\cite{wig}.
We shall review the construction here for completeness.
As usual, the Kac-Moody algebra defines a conformal family.
In analogy to the Sugawara construction, a set of Virasoro generators
can be formed which are bilinear in the elements $ISO(2,1)$ algebra.
These Virasoro generators were shown to have zero central charge
\cite{sss}.

In this article, we shall also apply the formalism of refs.
\cite{cs1,cs2}
to the case of Chern-Simons gravity on a disc with holes.
The holes correspond to sources for the curvature and torsion two forms.
In the limit where the holes
shrink to points, the points can be viewed of as massive spinning
point particles. As in previous treatments \cite{sta,gri,djt,ks,sss,ss},
the charges associated with these sources
are just the particle momenta and angular momenta.
Here however,
more information is needed to give a complete description of the system.
This is because to each hole we must assign an entire
$ISO(2,1)$ Kac-Moody algebra.
Thus each particle carries a unitary representation
of the infinite dimensional
algebra, and the Hilbert space consists of tensor products of
such unitary representations.  Unlike in traditional approaches,
we thereby are able to regularize the system,
taking into account self-interactions of the
particles in a natural manner.

It is known how to construct the Fubini-Veneziano vertex operator
in Chern-Simons theory \cite{cs2}.  This operators when acting on
a ``sourceless" state, creates a source with a discrete charge.
We construct
the analogue of the Fubini-Veneziano vertex operator for Chern-Simons
gravity.  \it When acting on a
``sourceless" state, it creates a particle with a quantized mass. \rm
The lowest mass that the
particle can have is precisely
that needed to make the space-time geometry around the particle
that of a cylinder $\times$ ${\bf R}^{1}$.  The possibility of
such a space-time geometry was mentioned in ref. \cite{djt}.
The $n^{th}$ state has a mass of $n$-times the lowest mass, and it yields
an $n$-fold covering of the cylinder.

The plan of this article is as follows:  In Section 2, we apply
the canonical methods of \cite{cs1} to Chern-Simons gravity on
a disc.  The essential ingredient is that the
Gauss law constraint be differentiable, as well first class (in the
sense of Dirac).  The next step is finding a complete set of
observables for the theory.  By definition, they must have zero
Poisson brackets with the first class constraints.  We shall show
that the algebra of observables is precisely the $ISO(2,1)$
Kac-Moody algebra of ref. \cite{sss}.  In Section 2, we also
construct the corresponding Virasoro generators.  Unitary
representations of the $ISO(2,1)$
Kac-Moody algebra, which describe the quantum theory for Chern-Simons
gravity on a disc, are reviewed in Section 3 \cite{sss}.
The quantum theory
 is extended to include sources in Section 4.  In Section 5,
we define the vertex operator for $2+1$ gravity, and we show how it
can create cylindrical geometries.

Section 6 represents a departure from the previous sections and
also from previous treatments of Chern-Simons theories on manifolds with
boundaries.  In those treatments, all the connection one forms are
unconstrained at the boundaries.
Then, in order to make the Gauss law constraint
differentiable, it is necessary to impose boundary
conditions on the test or smearing functions used in
defining the Gauss law constraint.  For Chern-Simons gravity, the
alternative possibility exists of imposing conditions on the
fields at the boundary, while relaxing the conditions on the
test functions.  In Section 6, we are able to classify all possible
such boundary conditions on the fields and test functions consistent with
the requirement that the Gauss law constraint
be differentiable and, also,
first class.  One suprising result is that in many instances, the
field content of the
(unconstrained) connection forms remaining on the boundary
are just those needed to define \it two dimensional \rm gravity.
We speculate that it may be possible to generate a variety of different
two dimensional gravity theories starting from three dimensional ones.

Concluding remarks are made in Section 7, including comments on
the spin-statistics theorem for particles in $2+1$ gravity.

\sxn{THE CANONICAL FORMALISM ON A DISC}

\indent
The $ISO(2,1)$ Chern-Simons
action on the solid cylinder $D \times {\bf R}^{1}$ is given by
\cite{wit}
\begin{equation}
S= \kappa \int_{D \times {\bf R}^1} e_a \wedge \Bigl(d\omega^a
+ {1\over 2}\epsilon^{abc}\omega_b\wedge\omega_c\bigr) \;,
\end{equation}
where $e^a$ and $\omega_a$, $a=0,1,2$, are dreibein one forms and
spin connection one forms, respectively.  Together,
$e^a$ and $\omega_a$  define the
$ISO(2,1)$ connection one forms.  The indices $a,b,c,...$ are raised and
lowered using the Minkowski metric, $\eta = {\rm diag}(-1,1,1)$.
$\epsilon^{abc}$ is the Levi-Civita symbol with
$\epsilon_{012}=1 \;.$
The constant $\kappa$ is related to the three dimensional
gravitational constant.

We now apply the canonical approach of ref. \cite{cs1,cs2} to
the action (2.1).  As time is indispensable in this approach,
we arbitrarily choose a time  function
denoted henceforth by $x^0$. Any constant $x^0$ slice of the solid
cylinder is then the disc $D$ with coordinates $x^1$, $x^2$.
The phase space of the action $S$ is spanned by
$\omega_i^a$ and $e_i^a,\; i,j=1,2$, the components of the one forms
$\omega^a$ and $e^a$, respectively, on the $x^0$ slice.
(The components $e^a_0$ and $\omega^a_0$ do not occur as
coordinates of phase space.  This is because
their conjugate momenta are weakly zero and first class, in the sense of
Dirac.)
$\omega_i^a$ and $e_i^a$ satisfy the equal time Poisson brackets (PB's):
$$ \left \{\omega_i^a(x),\omega_j^b(y)\right \}=
\left \{e_i^a(x),e_j^b(y)\right \}=   0,  $$
\be
\left \{\omega_i^a(x),e_j^b(y)\right \}={1\over{\kappa}}
\epsilon_{ij} \eta^{ab}\delta^2(x-y) \;,\;
\ee
as well as the Gauss law constraint.
Here, we define the two-indexed Levi-Civita symbol by
 $\epsilon_{ij}=\epsilon_{0ij};.$
To write the Gauss law constraint, we introduce
test or smearing functions on $D$.  We denote them by
$\Lambda^{(0)}= \{\Lambda_a^{(0)}(x),a=0,1,2\}$ and
$\Sigma^{(0)}= \{\Sigma_a^{(0)}(x),a=0,1,2\}$.  The Gauss law constraint
is given by:
\be
g(\Lambda^{(0)},\Sigma^{(0)})  \; =  {\kappa \over 2}
 \int_{D}\Bigl( \Lambda_a^{(0)}(x) R^a(x)
+ \Sigma_a^{(0)}(x) T^a(x) \Bigr) \approx 0 \; ,
\ee
where $R^a$ and $T^a$ are the $SO(2,1)$ curvature and the
torsion two forms, respectively,
\be
R^a=d\omega^a + {1\over 2}\epsilon^{abc}\omega_b\wedge \omega_c\; ,\quad
T^a=de^a + \epsilon^{abc}e_b\wedge \omega_c \;,
\ee
and $\approx$ denotes weak equality in the sense of Dirac.

It remains to state the space ${\cal T}^{(0)}$ of test functions,
$\Lambda^{(0)}$ and $\Sigma^{(0)}$.  According to ref. \cite{cs1,cs2},
we require that the Gauss law constraint is
differentiable, and also first class.  Differentiability requires
that the smearing functions be continuous on D, and that the
Gauss law can be relied upon
to generate well defined canonical transformations on phase space.
The Gauss law constraints
further generate the $ISO(2,1)$ gauge transformations, provided
they are first class.

By varying $g(\Lambda^{(0)},\Sigma^{(0)})$
with respect to $\omega^a$ and $e^a$, we obtain
$${2\over \kappa} \delta g (\Lambda^{(0)},\Sigma^{(0)}) =
\; -\;\int_{D}\Bigl( d \Lambda_a^{(0)}-
\epsilon_{abc}\Lambda^{(0)b} \omega^c -
\epsilon_{abc}\Sigma^{(0)b} e^c\Bigr) \wedge
 \delta \omega^a \qquad\qquad  $$
\be
\qquad-\;\int_{D}\Bigl( d \Sigma_a^{(0)}-
\epsilon_{abc}\Sigma^{(0)b} \omega^c\Bigr) \wedge
 \delta e^a \;+ \;
\int_{\partial D}\Bigl(\Lambda_a^{(0)} \delta \omega^a  +
\Sigma_a^{(0)} \delta e^a \Bigr)    \; .
\ee
By definition, $g(\Lambda^{(0)},\Sigma^{(0)})$
is differentiable in $e^a$ and $\omega^a$ only if the
boundary term,
$\int_{\partial D}\Bigl(\Lambda_a^{(0)} \delta \omega^a  +
\Sigma_a^{(0)} \delta e^a \Bigr)    \; ,$
 is zero.
If we do not wish to constrain the phase
space by legislating that $\delta e^a$ and $\delta \omega^a$
be zero on $\partial D$ to achieve
this goal, we are led to the following conditions
on the test functions $\Lambda^{(0)}$ and $\Sigma^{(0)}$
in ${\cal T}^{(0)}$:
\be
\Lambda^{(0)} \mid_{\partial D} = 0 \;, \quad
\Sigma^{(0)} \mid_{\partial D} = 0 \;.
\ee

With these conditions, the Poisson brackets of
$g(\Lambda^{(0)},\Sigma^{(0)})$ with $\omega^a$ and  $e^a$ are
given by
$$ \{g (\Lambda^{(0)},\Sigma^{(0)}),\omega^a_i(x)\}\;=\;
{1\over 2} \Bigl( \partial_i \Sigma^{(0)a}
 - \epsilon^{abc}\Sigma^{(0)}_b  \omega_{ci}\Bigr)(x) \; ,\qquad $$
\be
 \{g (\Lambda^{(0)},\Sigma^{(0)}),e^a_i(x)\}\;=\;
{1\over 2} \Bigl( \partial_i \Lambda^{(0)a}
 - \epsilon^{abc}\Lambda^{(0)}_b  \omega_{ci}
 - \epsilon^{abc}\Sigma^{(0)}_b e_{ci}\Bigr)(x) \;,
\ee
from which we can compute the Poisson brackets of the
$g(\Lambda^{(0)},\Sigma^{(0)})$'s
$$
 \{g(\Lambda^{(0)},\Sigma^{(0)}), g(\Lambda'^{(0)},\Sigma'^{(0)})\}\;=\;
g (\Lambda''^{(0)},\Sigma''^{(0)})\;+\;
{\kappa\over 4}
\int_{\partial D}\Bigl(\Sigma^{(0)a}d\Lambda'^{(0)}_a  +
\Lambda^{(0)a} d\Sigma'^{(0)}_a\Bigr)
$$
\be
+{\kappa \over 2}\int_{\partial D}\Bigl(\Sigma''^{(0)}_a e^a +
\Lambda''^{(0)}_a \omega^a  \Bigr)
\ee
where
$$ \qquad
\Lambda''^{(0)}_a = {1\over 2}\epsilon_{abc}\Bigl(\Sigma^{(0)b}
\Lambda'^{(0)c}
+\Lambda^{(0)b}\Sigma'^{(0)c}\Bigr) \;,$$
\be
 \Sigma''^{(0)}_a = {1\over 2}\epsilon_{abc}\Sigma^{(0)b}
 \Sigma'^{(0)c} \;.\qquad
\ee
The boundary terms in eq. (2.8) vanish
upon imposing the conditions (2.6) on all the test functions, and hence
$g(\Lambda^{(0)},\Sigma^{(0)})$ $\approx 0$ are first
class constraints.

Next we write down the observables of the theory.  By definition,
they have zero Poisson brackets with the first class constraints,
and hence are first class variables.  Here, they are of the form
\be
  q (\Lambda,\Sigma) =
\; -\;{\kappa\over 2}\int_{D}\biggl( d \Lambda_a\wedge\omega^a\;+\;
d\Sigma_a \wedge e^a   \; -  \;
{1\over 2}\epsilon_{abc}\Lambda^a\omega^b\wedge \omega^c \;- \;
\epsilon_{abc}\Sigma^a e^b\wedge \omega^c\biggr) \;,
\ee
where in this case, the test functions
$\Lambda\mid_{\partial D}$ and $\Sigma\mid_{\partial D}$
are not necessarily zero.  We shall identify $ q (\Lambda,\Sigma) $
with $ q (\Lambda',\Sigma') $ if the boundary values of $\Lambda$
and $\Sigma$ agree with those of $\Lambda'$ and $\Sigma'$, ie.
$\Lambda\mid_{\partial D} =\Lambda'\mid_{\partial D}$
and $\Sigma\mid_{\partial D}=\Sigma'\mid_{\partial D}$.  For then
$  q (\Lambda,\Sigma) -   q (\Lambda',\Sigma') =
 g (\Lambda-\Lambda',\Sigma-\Sigma') \approx 0\;.$
The $ q (\Lambda,\Sigma) $'s are
differentiable with respect to $\omega^a$ and $e^a$
for arbitrary test functions $\Lambda$ and $\Sigma$.  To see that
they correspond to first class variables,
we compute their Poisson brackets with
$g(\Lambda^{(0)},\Sigma^{(0)})$:
$$
 \{g (\Lambda^{(0)},\Sigma^{(0)}), q (\Lambda,\Sigma)\}\;=\;
g (\Lambda'^{(0)},\Sigma'^{(0)})\;+\;
{\kappa\over 4}\int_{\partial D}\Bigl(\Sigma^{(0)a}d\Lambda_a  +
\Lambda^{(0)a} d\Sigma_a\Bigr)$$
\be
\approx\;0\quad,\qquad
\ee
where
$$\qquad \Lambda'^{(0)}_a = {1\over 2}\epsilon_{abc}
\Bigl(\Sigma^{(0)b}\Lambda^c
+\Lambda^{(0)b}\Sigma^c\Bigr) \;,$$
\be
 \Sigma'^{(0)}_a = {1\over 2}\epsilon_{abc}\Sigma^{(0)b}\Sigma^c \;.
\qquad
\ee
The Poisson brackets of the $ q (\Lambda,\Sigma) $'s with themselves are
\be
 \{q (\Lambda,\Sigma), q (\Lambda',\Sigma')\}\;=\;
q (\Lambda'',\Sigma'')\;+\;
{\kappa\over 4}\int_{\partial D}\Bigl(\Sigma^a d\Lambda'_a  +
\Lambda^a d\Sigma'_a\Bigr) \;,
\ee
where
$$\qquad \Lambda''_a = {1\over 2}\epsilon_{abc}
\Bigl(\Sigma^b\Lambda'^c
+\Lambda^b\Sigma'^c\Bigr) \;,$$
\be
 \Sigma''_a = {1\over 2}\epsilon_{abc}\Sigma^b\Sigma'^c \;.
\qquad
\ee

It is known that the observables of Chern-Simons theory written
on a disc are elements of a Kac-Moody algebra associated with some
Kac-Moody group.  The Kac-Moody group is
an extension of the finite dimensional group $G$ in which the
gauge transformations of Chern-Simons theory are
defined.  Here $G=ISO(2,1)$,
so it follows that the observables $ q (\Lambda,\Sigma) $
must generate the $ISO(2,1)$ Kac-Moody group.
The brackets (2.13) define the $ISO(2,1)$ Kac-Moody algebra.

If we like, we can replace the elements $ q (\Lambda,\Sigma) $
of the algebra by another set of variables
$P^a (\psi)$ and $J^a(\psi)$ which were used in ref. \cite{sss}.
[These variables were interpreted, respectively,
 as momentum and angular momentum
current densities on the boundary $\partial D$.]
For this, we introduce the test functions
$\Xi^{(\psi,a)}= \Bigl\{\Xi^{(\psi,a)}_b(x)\Bigr\}$
whose values on the boundary $\partial D$ correspond to a delta function
centered around the point $\psi$ on $\partial D$.  More precisely,
let us introduce polar coordinates
$r$, $\theta$ on $D$ (with $r=r_0$ defining the boundary
$\partial D$, and $\theta$ parametrizing $\partial D$).
Then the boundary value of
$\Xi^{(\psi,a)}_b (r,\theta)$ is given by
$$\Xi^{(\psi,a)}_b (r_0,\theta)=
\delta_b^a\;\delta(\psi-\theta)\;. $$

We can now define $P^a (\psi)$ and $J^a(\psi)$ as follows:
\be
P^a(\psi) =2\;q(\Xi^{(\psi,a)},0)\quad {\rm and}\quad
J^a(\psi) =-2\;q (0,\Xi^{(\psi,a)})\; .
\ee

$P^a(\psi)$ and $J^a(\psi)$
can be used as a basis for the set of observables
$\{q (\Lambda,\Sigma)\}$.  To write down
 the completeness relation,
let us again utilize polar coordinates
$r$, $\theta$ on $D$ (with $r=r_0$ defining the boundary
$\partial D$).  Then
 \be
q (\Lambda,\Sigma)\approx  {1\over 2}
\int^{2\pi}_0 d\psi\Bigl(
\Lambda^a(r_0,\psi)P_a(\psi)
-\Sigma^a(r_0,\psi)J_a(\psi) \Bigr)
\ee
To prove this relation, let us write
$P^a(\psi)$ and $J^a(\psi)$ in polar coordinates:
\be
P^a(\psi) = -\kappa\;\omega^a_\theta(r_0,\psi)
+\kappa\int_D \Xi^{(\psi,a)}_b R^b      \;,
\ee
\be
J^a(\psi) = \kappa\; e^a_\theta(r_0,\psi)
-\kappa\int_D \Xi^{(\psi,a)}_b T^b     \;,
\ee
where $\omega^a=\omega^a_rdr+\omega^a_\theta d\theta$ and
$e^a=e^a_rdr+e^a_\theta d\theta$.  Then the right hand side of
eq. (2.16) can be written
$$
 -{\kappa\over 2}\int^{2\pi}_0 d \psi\Bigl(
 \Lambda_a(r_0,\psi)\omega^a_\theta(r_0,\psi)
+\Sigma_a(r_0,\psi)e^a_\theta(r_0,\psi) \Big)$$
\be
\;+\;{\kappa\over 2}\int^{2\pi}_0 d\psi
\int_D \Xi^{(\psi,a)}_b \Bigl(
\Lambda_a(r_0,\psi) R^b  +
\Sigma_a(r_0,\psi) T^b \Bigr)  \;,
\ee
while the left hand side of eq. (2.16) can be written
\be
 - {\kappa \over 2}
 \int_{\partial D}\Bigl( \Lambda_a \omega^a
+ \Sigma_a e^a(x) \Bigr)  \;+\;    {\kappa \over 2}
 \int_{D}\Bigl( \Lambda_a R^a + \Sigma_a T^a \Bigr) \;.
\ee
The difference of these two expressions is
\be
 {\kappa \over 2}  \int_{ D}\Biggl\{\Lambda_a - \int^{2\pi}_0 d\psi \;
 \Xi^{(\psi,b)}_a \Lambda_b(r_0,\psi)\Biggr\} R^a \; +   \;
 {\kappa \over 2} \int_{ D}\Biggl\{ \Sigma_a - \int^{2\pi}_0 d\psi \;
 \Xi^{(\psi,b)}_a \Sigma_b(r_0,\psi)\Biggr\} T^a \; .
\ee
Finally, we note that the boundary value of the functions in braces
$\{\}$ is zero.  These functions are therefore defined in the text
function space ${\cal T}^{(0)}$, and hence the expression (2.21)
corresponds to $g$ in the Gauss law constraint.
It must vanish (weakly), proving (2.16).

The Poisson bracket algebra of $P_a$ and $J_a$ is easily obtained
from eqs. (2.13) and (2.14).  We find:
\be
\{P_a(\psi),P_b(\psi')\} = 0         \;,
\ee
\be
\{J_a(\psi),J_b(\psi')\} = -\epsilon_{abc}J^c(\psi)
\delta(\psi-\psi')  \;,
\ee
\be
\{J_a(\psi),P_b(\psi')\}= -\epsilon_{abc}P^c(\psi)
\delta(\psi-\psi')
-\kappa \eta_{ab}\partial_\psi \delta(\psi-\psi')    \;.
\ee
This algebra is identical to that found in ref. \cite{sss}.
The second term on the left hand side of eq. (2.24) defines
the central extension to the Poincare loop group algebra.

We next construct the diffeomorphism generators $\ell$ of
$D$.  Since the action is diffeomorphism invariant, we know that
they are associated with first class constraints, and hence should
vanish weakly.  Also, as before, we shall require differentiability.
Following ref. \cite{cs1,cs2}, we can write
\be
\ell(v^{(0)}) = -\kappa\int_D v^{(0)i}(e^a_i R_a
+\omega^a_i T_a) \;,\quad
v^{(0)i}\mid_{\partial D} =  0\;,
\ee
which is weakly zero.  The restriction of the vector valued function
 $v^{(0)}$ on the boundary insures that $\ell(v^{(0)})$
is differentiable with respect to $e^a$ and $\omega^a$.  As a result,
$\ell(v^{(0)})$ generates transformations which vanish on $\partial D$.

Eq. (2.25) can be generalized
in order to obtain nontrivial transformations on the boundary.  For
this we replace $v^{(0)}$ by $v$ and set
\be
\ell(v) =-\kappa \int_D v^i(e^a_i R_a
+\omega^a_i T_a) \;+\;{\kappa\over 2}
\int_D d(v^i e^a_i \omega_a  \;+\;
v^i \omega^a_i e_a )   \;.
\ee
The requirement of differentiability can now be satisfied if
we impose the following boundary conditions for
the vector valued function $v^i$:
\be
 v^{i}|_{\partial D}(\theta)=
\epsilon(\theta)\left({\frac{\partial x^{i}}{\partial
\theta}}\right) {\Bigg|_{\partial D}}       \; ,
\ee
where $x_i$ denote the space coordinates,
 $\theta$ is an angular coordinate
parametrizing $\partial D$ and $\epsilon (\theta)$ is an arbitrary
function.  To show that $\ell(v)$ is first class, we can
take its Poisson bracket with the Gauss law constraint
\be
\left\{\ell(v ),g(\Lambda^{(0)},\Sigma^{(0)})\right\}
= g ({\cal L}_{v} \Lambda^{(0)},{\cal L}_{v} \Sigma^{(0)})
 \approx 0~\;,
\ee
where ${\cal L}_v \Lambda^{(0)}$ is the Lie derivative of the
scalar field
$ \Lambda^{(0)}$ with respect to $v$ and is defined by
${\cal L}_{v}\Lambda^{(0)} =  v^{j}\partial_{j}\Lambda^{(0)}\;.$
It also of course follows that $\ell(v^{(0)})$ is first class.
Similar Poisson brackets are obtained for
$\ell(v)$ with $q (\Lambda,\Sigma)$
\be
\left\{\ell(v ),q(\Lambda,\Sigma)\right\}
= q ({\cal L}_{v} \Lambda,{\cal L}_{v} \Sigma)\;,
\ee
while the Poisson bracket of two $\ell$'s gives the usual Virasoro
algebra
\be
\{\ell(v),\ell(v')\} =\ell({\cal L}_{v}v') \;.
\ee
Now ${\cal L}_{v}v'$ denotes
the Lie derivative of the vector field $v'$ with respect to the vector
field $v$ and is given by
$({\cal L}_{v}v')^{i} =  v^{j}\partial_{j}v'^{i}-
v'^{j}\partial_{j}v^{i}\;.$

As usual,
the Virasoro generators can be expressed (weakly) in terms of a
product of Kac-Moody generators.
The Sugawara construction for $ISO(2,1)$ is as follows:
\be
\ell(v)\approx
-{1\over\kappa}\int_{\partial D}d\psi\;\epsilon(\psi)
P_a(\psi)J^a(\psi) \;.
\ee
The proof of (2.31) is essentially the same as in ref. \cite{cs1}.
For this, we can again use polar coordinates
$r$, $\theta$ on $D$ (with $r=r_0$ defining the boundary
$\partial D$).
{}From the two expressions (2.15) for $P^a$ and $J^a$,
the right hand side of eq. (2.31) is weakly equal to
\be
\kappa\int_{\partial D}d\psi\;\epsilon(\psi) \Biggl\{
[\omega^a_\theta e_{a\theta}](r_0,\psi) - \int_D
\Xi^{(\psi,a)}_b \Bigl(R^b e_\theta^a(r_0,\psi)
+T^b \omega_\theta^a(r_0,\psi) \Bigr)\Biggr\}   \;.
\ee
In the above, we have dropped terms quadratic in
the curvature $R$ and the torsion $T$ due to the Gauss law
constraint.  In this regard, the relevant test function $\Sigma^{(0)}$
(or $\Lambda^{(0)}$) for the Gauss law constraint (2.3),
involves $R$ (or $T$) itself.

Concerning the left hand side of eq. (2.31), if we substitute the
boundary value of $v^i$ given in (2.27), into eq. (2.26) we get
\be
 \kappa\int_{\partial D}d\psi\;\epsilon(\psi)
[\omega^a_\theta e_{a\theta}](r_0,\psi)
-\kappa \int_D v^i(e^a_i R_a+\omega^a_i T_a) \; .
\ee
In comparing (2.32) with (2.33),
we find their difference to be equal to $$
\kappa \int_D\Biggl\{ v^i e^a_i - \int_{\partial D}d\psi\;\epsilon(\psi)
\Xi^{(\psi,b)a}  e_{\theta b}(r_0,\psi)\Biggr\}R_a  $$
\be
+\kappa \int_D\Biggl\{ v^i \omega^a_i -
\int_{\partial D}d\psi\;\epsilon(\psi)
\Xi^{(\psi,b)a}  \omega_{\theta b}(r_0,\psi)\Biggr\}T_a  \;.
\ee
Finally,
 we note that the boundary value of the functions in braces $\{\}$
is zero.  These functions are therefore defined in the space
${\cal T}^{(0)}$, and hence the expression (2.34) corresponds to $g$
the Gauss law constraint.  It must vanish (weakly), proving (2.31).

\sxn{THE QUANTUM THEORY}

To quantize the system described in the previous section, it is
sufficient to find unitary representations of the $ISO(2,1)$ Kac-Moody
algebra.  This was already done in ref. \cite{sss}.
We review it here for completeness.

In the previous section, we found that
$P_a (\psi)$ and $J_a(\psi)$ formed a basis for the
algebra.  To quantize, we replace the phase space variables
$P_a (\psi)$ and $J_a(\psi)$
by quantum operators ${\bf P}_a (\psi)$ and ${\bf J}_a(\psi)$.
We replace Poisson brackets (2.22-24)
by -$i$ times the commutator bracket,
thereby obtaining the quantum version of the Kac-Moody algebra:
\be
[{\bf P}_a(\psi),{\bf P}_b(\psi')] = 0         \;,
\ee
\be
[{\bf J}_a(\psi),{\bf J}_b(\psi') ]  = -i\epsilon_{abc}{\bf J}^c(\psi)
\delta(\psi-\psi')  \;,
\ee
\be
[{\bf J}_a(\psi),{\bf P}_b(\psi') ]= -i\epsilon_{abc}{\bf P}^c(\psi)
\delta(\psi-\psi')
-\kappa \eta_{ab}\partial_\psi \delta(\psi-\psi')    \;.
\ee
${\bf P}_a (\psi)$ and ${\bf J}_a(\psi)$ commute with
${\bf g}(\Lambda^{(0)},\Sigma^{(0)})$,
the quantum analogue of $g(\Lambda^{(0)},\Sigma^{(0)})$.
$ISO(2,1)$ gauge invariance in the quantum theory follows by
 requiring that  ${\bf g}(\Lambda^{(0)},\Sigma^{(0)})$ annihilates
the states of the Hilbert space.

The Hilbert space consists of unitary representations of the Kac-Moody
algebra eqs. (3.1-3)
Highest weight constructions are standardly employed for the purpose
of finding unitary representations of Kac-Moody algebras \cite{go}.
However, this procedure is complicated, at best,
if the underlying group generated by the charges
is noncompact.  Here, the underlying group generated by the charges
$\int^{2\pi}_0 d\psi {\bf P}_a(\psi)$ and
$\int^{2\pi}_0 d\psi {\bf J}_a(\psi)$
is $ISO(2,1)$.  It is
not only noncompact, but also a semidirect product group.
A highest weight construction for this algebra was attempted in ref.
\cite{sss},
but it failed to give a nontrivial representation of the operators
${\bf P}_a (\psi)$ and ${\bf J}_a(\psi)$.

On the other hand, an alternative approach was given in ref. \cite{sss}
which did yield nontrivial unitary representations for
the algebra of ${\bf P}_a (\psi)$ and ${\bf J}_a(\psi)$.
It employed the method of induced representations \cite{wig},
which is commonly used for finding unitary representations of
semidirect product groups.  We describe it below.

The method utilizes the fact that the momentum density operators
${\bf P}_a (\psi)$ commute and hence are simultaneously diagonalizable.
States in the Hilbert space
can therefore be labeled by the eigenvalues
$ P_a (\psi)$ of ${\bf P}_a (\psi)$.
To construct a certain unitary representation, we first pick a
 ``standard" set of eigenvalues, which we denote by
$\hat{P}=\{ \hat{P}_a (\psi)\}$, belonging to that representation.
Let $|\hat{P},1,\alpha>$ be its corresponding eigenvector.  Thus
\be
{\bf P}_a (\psi)  | \hat{P},1,\alpha> =
\hat{ P}_a (\psi)  | \hat{P},1,\alpha> \;.
\ee

We will see that
the states of a given representation are
not completely labeled by the eigenvalues of
${\bf P}_a (\psi)$, hence the need for
a degeneracy index $\alpha$ in the state vector
$ | \hat{P},1,\alpha> $.

In order to obtain the remaining states of the representation,
let us introduce the matrix $M=M_{ab}(\psi)$, which
denotes a local Lorentz transformation,
and further,
let ${\bf U}(M)$ be a unitary representation of $M$.  We want to
express the latter as an exponential of the angular momentum operators
${\bf J}_a (\psi)$.  For
then, the commutation relation (3.1-3)
leads to the following operator equation
\be
{\bf U}(M)^{-1}{\bf P}_a(\psi) {\bf U}(M)=M_{ab}(\psi){\bf P}^b(\psi)
-{\kappa\over 2}\epsilon_{abc}[ \partial_\psi M M^{-1} ]^{bc}(\psi)\;.
\ee
The first term in on the right-hand-side of eq. (3.5) represents a
local Lorentz transformation of the momentum operator, while the
second term results from the central term in the commutation relation
(3.3).

Now the remaining states of the
representation are obtained by having ${\bf U}(M)$
act on the ``standard state" $ | \hat{P},1,\alpha> $.  From eq. (3.5),
the momentum density eigenvalues of the resulting states will be of
the form
\be
P_a(\psi)= M_{ab}(\psi)\hat{ P}^b(\psi)
-{\kappa\over 2}\epsilon_{abc}[ \partial_\psi M M^{-1} ]^{bc}(\psi)\;.
\ee
The set of all such $P=\{P_a(\psi)\}$ defines an orbit in the space of
momentum densities.  We can label the orbit by the point
$\hat{P}=\{\hat{P}_a(\psi)\}$
in momentum density space through which it passes.

A state with a given momentum density eigenvalue $P$ on any particular
orbit is defined only up to the
action of the little group $G_P$
 of $P$, ie. the subset of local Lorentz
transformations that leaves the momentum density eigenvalue unchanged.
The little groups $G_P$ for all points $P$ on any particular orbit
are isomorphic.
Let $G_{\hat{P}}=\{\hat{M}\}$ be the little group associated with
the ``standard" eigenvalue $\hat{P}$.  Then from eq. (3.6), $\hat{M}$
satisfies
\be
\hat{P}_a(\psi)=  \hat{M}_{ab}(\psi) \hat{P}^b(\psi)
-{\kappa\over 2}\epsilon_{abc}[ \partial_\psi
\hat{M} \hat{M}^{-1} ]^{bc}(\psi)\;.
\ee
When ${\bf U}(\hat{M})$ acts on $ | \hat{P},1,\alpha> $ it can only
change the degeneracy index $\alpha$. Thus
\be
{\bf U}(\hat{M}) | \hat{P},1,\alpha> =
D_{\alpha\beta}(\hat{M})   | \hat{P},1,\beta>  \;,
\ee
where we define $D_{\alpha\beta}(\hat{M})$ to be some
 unitary representation
of the little group $G_{\hat{P}}$.

In ref. \cite{sss},
it was shown that $G_{\hat{P}}$ is a finite dimensional group.
We shall not repeat the proof here.
More specifically, $G_{\hat{P}}$ is isomorphic to either
$SO(2,1)$, $U(1)$ or ${\bf R}^1$.  The group $SO(2,1)$
results, for instance, when we set the standard momentum densities
$ \hat{P}^b(\psi)=0$.  In that case, eq. (3.7) shows that
$\{\hat{M}\}$  is the set of all $\psi$-independent $SO(2,1)$ group
matrices.
[Note in this case that the unitary representations $\{D_{\alpha\beta}\}$
are either trivial or infinite dimensional.]
$U(1)$ results when we set $ \hat{P}^b(\psi)=\delta^b_0\;\times\;const$.
Then $\{\hat{M}\}$ is the $SO(2)$ subset of
$\psi$-independent $SO(2,1)$ matrices.
When we set $ \hat{P}^b(\psi)=\delta^b_2\;\times\;const$,
$\{\hat{M}\}$ is an $SO(1,1)$ subset of
$\psi$-independent $SO(2,1)$ matrices, and hence isomorphic to
${\bf R}^1$.

In order to define a $ unique$ state with momentum density eigenvalues
$P=\{P^b(\psi)\}$ on an orbit through
$\hat{P}=\{\hat{P}^b(\psi)\}$ which are not equal
to the ``standard" eigenvalues, i.e.
$P\ne\hat{P}$, let us define a $ unique$ transformation
$M_P=M_P(\psi)$ which takes $ \hat{P}$ to $P$.
Thus $P_a= [ M_P ] _{ab}\hat{ P}^b
-{\kappa\over 2}\epsilon_{abc}[\partial_\psi M_P M_P^{-1}]
^{bc}$.
[For the above
example where $ \hat{P}^b(\psi)=0$, yielding the little group
$G_{\hat{P}}=SO(2,1)$, we can choose $M_P(\psi)$
such that it is the $SO(2,1)$ identity element at $\psi=0$.]
The unique state with momentum density eigenvalues $P^b(\psi)$,
which we denote by $| \hat{P},M_P,\alpha>  $,
can then be defined according to
\be
 | \hat{P},M_P,\alpha>  = {\bf U}(M_P) | \hat{P},1,\alpha> \;.
\ee
Then ${\bf P}_a (\psi)  | \hat{P},M_P,\alpha> =
 P_a (\psi)  | \hat{P},M_P,\alpha> $.

Now an arbitrary $M$ which takes
$ \hat{P}$ to $P$ can be written $M=M_P\hat{M}$, where
$\hat{M}\in G_{\hat{P}}$.
When ${\bf U}(M)$ acts on the state $ | \hat{P},1,\alpha> $, we get
\be
 {\bf U}(M) | \hat{P},1,\alpha>  =
 {\bf U}(M_P){\bf U}(\hat{M}) | \hat{P},1,\alpha>  =
D_{\alpha\beta}(\hat{M})   | \hat{P},M_P,\beta>  \;.
\ee

It remains to determine the action of an arbitrary
${\bf U}(N)$ on an arbitrary state $ | \hat{P},M_P,\alpha> $.
For this we define $\hat{N}=M^{-1}_{P'}N M_P$ and
$P'_a= N_{ab} {P}^b
-{\kappa\over 2}\epsilon_{abc}[ \partial_\psi N {N}^{-1}]^{bc}$.
It follows that $\hat{N}\in G_{\hat{P}}$ and
$$  {\bf U}(N) | \hat{P},M_P,\alpha>  =
 {\bf U}(M_{P'}){\bf U}(\hat{N}){\bf U}(M_P)^{-1} | \hat{P},M_P,\alpha>$$
$$={\bf U}(M_{P'}){\bf U}(\hat{N}) | \hat{P},1,\alpha>$$
\be
=D_{\alpha\beta}(\hat{N})   | \hat{P},M_{P'},\beta>  \;,
\ee
where we have used eq's (3.8) and (3.9).

To summarize, in analogy with the unitary
representations of the Poincare group,
unitary representations for $ISO(2,1)$ Kac-Moody group
can be specified by their orbits
in momentum density space, along with representation
$\{D_{\alpha\beta}\}$ of the little group $G_{\hat{P}}$.

\sxn{POINT SOURCES ON THE DISC}

Here we consider introducing point sources on the disc.  They are
characterized, in general, by their momenta and angular momenta,
which play the role of charges in the Chern-Simons theory.  $ISO(2,1)$
Chern-Simons theory with sources has
been been treated by many authors. \cite{ks,gri,ss,car,cap}
  We shall offer a new approach
below.

The equations of motion in the absence of any sources imply that
the curvature and torsion two forms, $R^a$ and $T^a$, vanish everywhere.
On the other hand, in the presence of a point source with the
space-time coordinates $z_\mu=z_\mu(\tau)$, where $\mu=0,1,2$ and
$\tau$ parametrizes the particle world line, this result is
modified to
\be
{\kappa\over 2} \epsilon^{\mu\nu\lambda} R_{\nu\lambda}^a (x)=
\int d\tau\;\delta^3 (x-z(\tau)) p^a \dot z^\mu \;,
\ee
\be
{\kappa\over 2} \epsilon^{\mu\nu\lambda} T_{\nu\lambda}^a (x)=
\int d\tau\;\delta^3 (x-z(\tau)) j^a \dot z^\mu \;.
\ee
Here $p^a=p^a(\tau)$ and
$j^a=j^a(\tau)$ are the particle momenta and angular momenta,
respectively, and $ R_{\nu\lambda}^a (x)$ and
$ T_{\nu\lambda}^a (x)$ are the space-time components of $R^a$
and $T^a$.  (For the moment we are not considering the effect of the
disc boundary $\partial D$.)

Equations of motion for the particle degrees of freedom $p^a$ and $j^a$,
are deduced from the Bianchi identities for the fields
\be
dR^a + \epsilon^{abc}\omega_b\wedge R_c =0\;,
\ee
\be
dT^a + \epsilon^{abc}\biggl(\omega_b\wedge T_c+e_b\wedge R_c\biggr)
 =0\;.
\ee
Upon substituting (4.1) and (4.2) into (4.3) and (4.4), we then get the
following equations for $p^a$ and $j^a$,
\be
\dot p_a +\epsilon_{abc}\omega^b_\mu(z) p^c \dot z^\mu =0\;,
\ee
\be
\dot j_a +\epsilon_{abc}\biggl(\omega^b_\mu(z) j^c  +
e^b_\mu(z) p^c\biggr) \dot z^\mu  =0\;.
\ee

These equations can also be derived starting from an action principal
\cite{ss,ks}.
In generalizing to
the case of more than one point particle, each particle would
satisfy equations of motion analogous to (4.5) and (4.6).

$\omega^b_\mu(z)$ and $e^b_\mu(z)$ in equations (4.5) and (4.6)
are components of the spin
connections $\omega^b$ and dreibein one forms $e^b$, evaluated at
the particle space-time position $z_\mu$.  If the particle is
treated as a test particle these functions are well defined.
However, in general, we must
consider self-interactions of the particle. In that case,
$\omega^b_\mu(z)$ and $e^b_\mu(z)$ are singular functions, ie.
$\omega^b=\omega^b_\mu (x)dx^{\mu}$ and
$e^b=e^b_\mu (x)dx^{\mu}$ have no definite limit
when $x$ approaches $z$.
This singularity demands regularization.

Following ref. \cite{cs2}, a good way to regularize is to
punch a hole $H$ containing $z$, and eventually to shrink the hole to
the point $z$.
Once this hole is made, the action is no longer defined
on a disc $D$, but on $D \setminus H$,
a disc with a hole. $D \setminus H$
has a new boundary $\partial H$ and it must be treated
exactly like $\partial D$.  Thus for instance, the
Gauss law must be changed to
\be
g(\Lambda^{(1)},\Sigma^{(1)})
 \approx 0 \; ,
\ee
where the new test functions $\Lambda^{(1)}$ and $\Sigma^{(1)}$
satisfy
\be
\Lambda^{(1)} \mid_{\partial D}=\Lambda^{(1)} \mid_{\partial H}
= 0 \;, \quad
\Sigma^{(1)} \mid_{\partial D} = \Sigma^{(1)} \mid_{\partial H} =
 0 \;.
\ee

There are now two $ISO(2,1)$ Kac-Moody algebras in the theory,
one each for each boundary, $\partial D$ and $\partial H$.
The elements of the algebras consist of
 the observables $q(\Lambda,\Sigma)$.  We can again use them to define
momentum and angular momentum densities
$P^{(A)a}(\psi)$ and $J^{(A)a}(\psi),\; A=0,1$.  $A=0$ corresponds
to currents on $\partial D$, while
$A=1$ corresponds to currents on $\partial H$.  We define these currents
in an analogous fashion to eq. (2.15).
For this we introduce test functions
$\Xi^{(\psi,a,A)}= \Bigl\{\Xi^{(\psi,a,A)}_b(x)\Bigr\}$, such that
$$\Xi^{(\psi,a,0)}_b (\theta)= \left\{
\matrix{\delta_b^a\delta(\psi-\theta), \quad {\rm on} \;\partial D \cr
0, \quad{\rm on }\;\partial H\cr} \right.\;, $$
and
$$\Xi^{(\psi,a,1)}_b (\theta)= \left\{
\matrix{0, \quad {\rm on }\;\partial D\cr
\delta_b^a\delta(\psi-2\pi+\theta), \quad
{\rm on} \;\partial H\cr }\right.\;. $$
As in Section 2,
$\theta$ is to be interpreted as an angular coordinate.
[The coordinates $\theta$ on both
$\partial D$ and $\partial H$ increase, say,
in the anticlockwise sense.  Both $\theta$ and $\psi$ are assumed to run
from $0$ to $2\pi$.]  We may now define $P^{(A)a}(\psi)$
and $J^{(A)a}(\psi)$ as follows:
\be
P^{(A)a}(\psi) =2\;q(\Xi^{(\psi,a,A)},0)\quad {\rm and}\quad
J^{(A)a}(\psi)=-2\;q (0,\Xi^{(\psi,a,A)})\; .
\ee

In the unregularized theory the particle dynamics is given in
terms of charges $p^a$ and $j^a$, while here it is described by
$P^{(A)a}(\psi)$ and $J^{(A)a}(\psi)$.  In the process of
regularizing the
theory we have effectively enlarged the phase space of the
particle.  Since $p^a$ and $j^a$ correspond the the total momenta
and angular momenta, these quantities should be identified with
$\int_0^{2\pi}d\psi\; P^{(A)a}(\psi)$ and
$\int_0^{2\pi}d\psi\;J^{(A)a}(\psi)$,
respectively, in the regularized theory.

In the quantum theory, we replace $P^{(A)a}(\psi)$
and $J^{(A)a}(\psi)$ by quantum operators
 ${\bf P}^{(A)a}(\psi)$ and ${\bf J}^{(A)a}(\psi)$.  The commutators
of these operators define the direct sum of two $ISO(2,1)$
Kac-Moody algebras.  The Hilbert space is obtained by taking
tensor products of two induced representations of the type
discussed in Section 3.   Unitary representations of the
Hilbert space are now characterized
by a pair of orbits in momentum density space.  The two orbits,
which we can label $(0)$ and $(1)$, pass through some ``standard"
points, $ \hat{P}^{(0)}  $ and $ \hat{P}^{(1)}  $.
We can then define a ``standard" eigenstate
$| \hat{P}^{(0)},1,\alpha>\otimes  $
 $| \hat{P}^{(1)},1,\beta>  $, associated with the ``standard"
eigenvalues $ \hat{P}^{(0)}  $ and $ \hat{P}^{(1)}  $.
In addition to the orbits in momentum density space, the tensor product
representations are labeled by unitary representations of the
little groups of $ \hat{P}^{(0)}  $ and $ \hat{P}^{(1)}  $.  Since
$\alpha$ and $\beta$ are indices associated with different such unitary
representations, and also, in general, different little groups, they
may range over different values.

 The remaining states of the Hilbert space are gotten by acting on
 the standard state $| \hat{P}^{(0)},1,\alpha>\otimes  $
 $| \hat{P}^{(1)},1,\beta>  $ with local Lorentz transformations
${\bf U}^{(0)}(M_{P^{(0)}})$ and ${\bf U}^{(1)}(M_{P^{(1)}})$
generated, respectively, by ${\bf J}^{(0)a}(\psi)$ and
${\bf J}^{(1)a}(\psi)$.  A general tensor product state
diagonal in ${\bf P}^{(A)}$ is thus
$$
 | \hat{P}^{(0)},M_{P^{(0)}},\alpha>  \otimes
 | \hat{P}^{(1)},M_{P^{(1)}},\beta> \qquad\qquad$$
\be
\qquad\qquad  = {\bf U}^{(0)}(M_{P^{(0)}}) \;
 {\bf U}^{(1)}(M_{P^{(1)}})  \;
   | \hat{P}^{(0)},1,\alpha> \otimes
   | \hat{P}^{(1)},1,\beta> \;,
\ee
with $P^{(A)}$ being the eigenvalues of ${\bf P}^{(A)}$.

It is straightforward to generalize these results to the case of $N$
sources or holes.  In that case, there would $N+1$ $ISO(2,1)$ Kac-Moody
algebras, one for each hole and one for the boundary $\partial D$.
The Hilbert space for the quantum theory is then obtained by taking
tensor products of $N+1$ induced representations of the type discussed
in Section 3.

\sxn{VERTEX OPERATORS}

Vertex operators of conformal field theory have been utilized for
the purpose of creating sources in Chern-Simons theories.  Here
we write down the analogue of the vertex operator for
topological gravity.  That operator is seen to create a source for the
curvature two form $R^a$.  The source can be interpreted as a massive
point particle in its rest frame.  The mass of the point particle
determines the surounding space-time metric.  The vertex operator for
topological gravity therefore creates certain space-time geometries.

To proceed we shall closely follow ref. \cite{cs2}.
Unlike ref. \cite{cs2}, however, we do not deal with highest weight
representations, but rather with the induced representations
discussed previously.

As in the previous section, we shall examine the manifold
$D\setminus H$ of a disc with a hole.
We eventually would like to take the limit where the hole is shrunk
to a point $z$.
The hole or point
is originally to be ``sourceless" (with regards to both the
$SO(2,1)$ curvature $R^a$ and the torsion $T^a$).
It should then be described
by a state $|0>$ where the momentum density associated with $\partial H$
vanishes, i.e. the eigenvalues of ${\bf P}^{(1)a}(\psi)$ are zero.
Let us call this the ``standard" state.  Hence $ \hat{P}^{(1)}=0$, and
\be
|0> =  | \hat{P}^{(0)},1,\alpha> \otimes  | 0,1,\beta> \;.
\ee
$ \hat{P}^{(0)}$ is the momentum density associated with
the disc boundary $\partial D$, which we can assume, in general,
to be nonzero.
$\beta$ labels unitary representations of the little group
$G_{\hat{P}^{(1)}=0}$, which is isomorphic to $SO(2,1)$.
Here we shall assume the trivial representations for
$G_{\hat{P}^{(1)}=0}$,
and thus $\beta$ is a trivial index which we set to zero.
In this case, from eq. (3.8) we have
\be
{\bf U}^{(1)}(\hat{M}) |0> =   |0>  \;.
\ee
${\bf U}^{(1)}(\hat{M}) $ is the unitary representation of
$\hat{M}\in  G_{\hat{P}^{(1)}=0}$.  (Our notation agrees with that in
the previous section, where ${\bf U}^{(1)}$
acts on the second ket in the tensor product.)
${\bf U}^{(1)}(\hat{M}) $ is generated by the angular
momentum operators, $\int_0^{2\pi}d\psi\; {\bf J}^{(1)a}(\psi)$,
associated with the hole.  From eq. (5.2), we
may then conclude that the angular momentum, as well as the
momentum, are zero for the state $|0>$.  In the limit where the hole
is shrunk to a point $z$, the point has no mass or spin.
Therefore, as desired, it is indeed ``sourceless" with regards
to both $R^a$ and $T^a$.

{}From ref. \cite{cs2}, the vertex operator can be written in terms
of $q$'s (defined in Section 2), whose test functions are linear
in the polar angle $\theta$.  $\theta$, and hence the vertex operator,
 is well defined on
$D\setminus (H \bigcup L_0)$, where the line $L_0$ from $\partial D$
to $\partial H$ has zero polar angle.
Accordingly, we define the quantity $\bar q=
  q (0,\bar\Sigma)$, where
\be
  \bar\Sigma_a =2\theta\delta_a^0 \;.
\ee
As we shall see shortly, this particular choice for the test functions
satisfies a rather restrictive criterion that the vertex operator
preserves the gauge invariance of states.

{}From eq. (5.3),
\be
  \bar q =  \; -\;\kappa\int_{D}\biggl( d\theta \wedge e^0   \; -  \;
\epsilon^{0bc}\theta e_b\wedge \omega_c\biggr) \;.
\ee
Its Poisson brackets with the observables $q(\Lambda,\Sigma)$ are
\be
 \{q (\Lambda,\Sigma),\bar q \}\;=\;
q (\bar{\bar\Lambda},\bar{\bar\Sigma})\;+\;
{\kappa\over 2}\int d\theta\Bigl(
\Lambda^0|_{\partial D}
-\Lambda^0|_{\partial H}\Bigr)
\;,
\ee
where
$$ \bar{\bar\Lambda}^a = \epsilon^{ab0}  \;
\theta\Lambda_b \;,\qquad
 \bar{\bar\Sigma}^a = \epsilon^{ab0}    \;
\theta\Sigma_b \;.$$
In quantum theory, this leads to the following
commutation relations with
 ${\bf P}^{(A)a}(\psi)$ and ${\bf J}^{(A)a}(\psi)$:
\be
[{\bf P}^{(0)}_a(\psi),\bar{q}] = i\epsilon_{ab0}\;
\psi  {\bf P}^{(0)b}(\psi) +i\kappa\delta^{0}_a   \;,
\ee
\be
[{\bf P}^{(1)}_a(\psi),\bar{q}] = i\epsilon_{ab0}\;
\psi  {\bf P}^{(1)b}(\psi) -i\kappa\delta^{0}_a   \;,
\ee
\be
[{\bf J}^{(0)}_a(\psi),\bar{q}] = i\epsilon_{ab0}  \;
\psi  {\bf J}^{(0)b}(\psi)  \;,
\ee
\be
[{\bf J}^{(1)}_a(\psi),\bar{q}] = i\epsilon_{ab0} \;
\psi  {\bf J}^{(1)b}(\psi)  \;.
\ee

We now define the vertex operator $V$ as follows:
\be
V=e^{in\bar{q}}  \quad.
\ee
{}From eq. (5.7),
it creates a state $V|0>$ which has a uniform ``energy" density
$P^{(1)}_0(\psi)=n\kappa$ at the hole boundary $\partial H$, i.e.
\be
{\bf P}^{(1)}_0(\psi)\;V|0>=n\kappa \;V|0>\quad.
\ee
The eigenvalues for the spatial-momentum density
${\bf P}^{(1)}_1(\psi)$ and ${\bf P}^{(1)}_2(\psi)$ are zero for $V|0>$.
In the limit where $H$ is shrunk to the point $z$, the state $V|0>$
describes a point particle located at $z$ with three-momenta
 \be
 p_a=2\pi n \kappa \delta^0_a \;.
 \ee
It therefore describes a particle with mass $m=2\pi n \kappa$
in its rest frame.
We shall see that the requirement that the state $V|0>$ be gauge
invariant, in addition to fixing the form (5.3) of
$ \bar\Sigma_a $, forces $n$ to be an integer.
It then follows that the vertex operator
can only create particles with a quantized mass spectrum, i.e.
\be
m= 2\pi\kappa\;\times\;{\rm integer}
\ee

The states $|0>$ and $V|0>$ belong to two different representations
of the Kac-Moody algebras.
This is evident because there exists no unitary transformation
$ {\bf U}^{(0)}(M^{(0)})\; {\bf U}^{(1)}(M^{(1)})$
[where $M^{(A)}$ are defined in the loop group of $SO(2,1)$],
connecting the two states, and also because the little
groups $G_{P^{(1)}}$ associated with the two states are different.
$G_{P^{(1)}=0}$ is the little group for $|0>$, while
$G_{P_a^{(1)}=n\kappa\delta^0_a}$ is the little group for $V|0>$.
The former is isomorphic to $SO(2,1)$, and the latter
 is isomorphic to $SO(2)$.  Further,
$G_{P_a^{(1)}=n\kappa\delta^0_a} \subset G_{P^{(1)}=0}$.

As mentioned earlier,
the state $|0>$ has zero angular momentum (associated with the hole).
The same is true for the state $V|0>$.  To prove this,
let $\hat{M}_0$ be an element of the little group
$G_{P_a^{(1)}=n\kappa\delta^0_a}$.  Then it is also
an element of the little group $G_{P^{(1)}=0}$.  From eq. (5.2),
${\bf U}^{(1)}(\hat{M}_0)$ acts trivially on $|0>$.
{}From eq. (5.9), ${\bf U}^{(1)}(\hat{M}_0)$
 commutes with the vertex operator, since
it is generated by the zero component
of the angular momentum operator, $\int_0^{2\pi}d\psi\;
{\bf J}^{(1)0}(\psi)$.
 Hence,
\be
{\bf U}^{(1)}(\hat{M}_0) \;V|0> =V \; {\bf U}^{(1)}(\hat{M}_0)|0> =V|0>
\ee
The last equality follows from eq. (5.2).  Thus upon assuming
trivial representation of the little group for $|0>$, we end
up with the trivial representation of the little group for $V|0>$.
Therefore the rotation generator,
i.e. zero component of angular momentum, is zero when acting on $V|0>$.
Since the corresponding point particle is in its rest frame, we can
conclude that it has no spin.  Although the
particle is a source of curvature, it is not a source for the torsion.

What space-time geometry
is created through the action of the vertex operator?
That is, what is the space-time metric associated with the point particle
with three-momenta (5.12)?

For this,
let us solve for the spin connection $\omega^a$ and the dreibein one form
$e^a$, starting from the field equations (4.1) and (4.2).
We can place the point particle at the origin of the coordinate system.
Hence, $z^1=z^2 = 0$.  For time, we set $z^0=x^0=\tau$.
Substituting eq. (5.12) into the field eq. (4.1), we find that the only
non-zero component of the curvature is
\be
R^0_{12}(x) = 2\pi n \delta^2(x)  \;.
\ee

A solution to (5.15) for the spin connections is just
\be
\omega^0 = nd\theta \;,
\ee
with $\omega^1=\omega^2=0$.  We can solve for the dreibein one forms
by setting the torsion equal to zero.  (This is since the particle
has no spin.)  A solution consistent with (5.16) is
\be
e^1={1\over r}\cos n \theta\; dr -n\sin n\theta\; d\theta\;,
\ee
\be
e^2={1\over r}\sin n\theta\; dr +n\cos n\theta \;d\theta \;.
\ee
The space-space components of the metric tensor now easily follow.
We find
\be
d\ell^2 ={1\over{r^2}}dr^2 +n^2 d\theta^2 \;.
\ee

For $n=1$, (5.19) is the invariant length in cylindrical space,
with $\ln r$ and $\theta$ being the coordinates of the cylinder
\cite{djt}.  Locally, a
cylindrical space is recovered as well for $n=2,3,...$.
This is clear if we replace $n \theta$ by a new angle $\theta'$.
Globally, we get an $n$-fold covering of the cylinder.

In the preceding analysis, we have ignored the boundary of the disc
$\partial D$.  From eq. (5.6),
the ``energy" density eignevalue $P^{(0)}_0(\psi)$ associated with the
disc boundary $\partial D$, has a contribution which is \it negative \rm,
\be
{\bf P}^{(0)}_0(\psi)\;V|0>=\biggr(
 \hat{P}^{(0)}_0 -n\kappa\biggl)\; V|0>\quad.
\ee
It thus appears that an antiparticle gets created at
$\partial D$.

We now examine the question of gauge invariance, and explain why
$n$ must be an integer.
Let us assume that the state $|0>$ is gauge invariant.
Here for a state to be gauge invariant it must be annihilated by
${\bf g}(\Lambda^{(1)},\Sigma^{(1)})$, the quantum analogue of
$ g(\Lambda^{(1)},\Sigma^{(1)})$, where the test functions
$\Lambda^{(1)}$ and $\Sigma^{(1)}$ are well defined on
$D \setminus H$ and satisfy eqs. (4.8).  To check whether
the state $V|0>$ is gauge invariant, we first note the
following property,
\be
V^{-1}{\bf g}(\Lambda^{(1)},\Sigma^{(1)})V =
{\bf g}(R_n\Lambda^{(1)},R_n\Sigma^{(1)}) \;,
\ee
where $R_n$ is an $SO(2)$ rotation matrix, which in polar
coordinates is given by
\be
R_n(\theta) =\pmatrix{
1 & 0 & 0 \cr
0 & \cos n\theta  & -\sin n\theta \cr
0 & \sin n\theta  & \cos n\theta \cr
} \quad.
\ee
Now upon redefining the test functions
$\Lambda^{(1)}$ and $\Sigma^{(1)}$, we have
\be
{\bf g}(\Lambda^{(1)},\Sigma^{(1)}) \;V|0> \;= V\;
{\bf g}(R_n^{-1}\Lambda^{(1)},R_n^{-1}\Sigma^{(1)}) |0>
\ee
The right-hand-side of eq. (5.23) is zero only if
$R_n^{-1}\Lambda^{(1)}$ and $R^{-1}_n\Sigma^{(1)}$
are well defined on $D \setminus H$, and satisfy eqs. (4.8).
Only then are they suitable test functions for the Gauss law on
$D \setminus H$.  But this happens only when
$R_n(\theta)$ is well defined on $D \setminus H$.
This necessarily implies that $n$ must be an integer,
thus leading to the quantized mass spectrum (5.13).

We see that although the test function $ \bar\Sigma_a $ used in defining
$\bar{q}$ is only defined on $D\setminus (H \bigcup L_0)$,
the test functions relevant for the Gauss law may be defined on
all of $D\setminus H$.  Requiring the latter puts a severe
restriction on the type of test functions we can use in defining
$ \bar q $, as well as on the values for $n$.
Basically, eq. (5.2) and
smooth deformations of this function (keeping
the values of $\bar\Sigma_a$ at $\theta=0$ and $2\pi$ fixed),
are the only possible such test functions.  (Actually, the values
 of $\bar\Sigma_a$ at the end points can be changed with
a corresponding change in $n$.)

In the above, we have examined the action of the vertex operator
acting on the ``sourceless" state $|0>$.
More generally, we can define the action of $V$ on any ``standard" state
$| \hat{P}^{(0)},1,\alpha>\otimes  $
 $| \hat{P}^{(1)},1,\beta>  $, according to
$$
V\; | \hat{P}^{(0)},1,\alpha>\otimes | \hat{P}^{(1)},1,\beta>\qquad\qquad
 $$
\be
\qquad\qquad  = {\bf U}^{(0)}(R_n) \;
 {\bf U}^{(1)}(R_n)  \;
   | \hat{P}^{(0)}_a-n\kappa\delta_a^0,1,\alpha> \otimes
   | \hat{P}^{(1)}_b+n\kappa\delta_b^0,1,\beta> \;,
\ee
Eq. (5.24) fixes any ordering ambiguities inherent the definition of $V$.
By acting on eq. (5.24) with unitary transformations
${\bf U}^{(A)}(M_{P^{(A)}})$ and using eq. (4.10), we then obtain
the action of $V$ on any state in the representation.

\sxn{ALTERNATIVE BOUNDARY CONDITIONS}

In Section 2, we chose the variations of
the fields $e^a$ and $\omega^a$ to be
unconstrained at the boundary $\partial D$ of a disc.
{}From eq. $(2.5)$, conditions then had to be imposed on the test
functions $\Lambda^{(0)}$ and $\Sigma^{(0)}$
in order for the Gauss law to be differentiable.
Specifically, the test functions for the Gauss law had to vanish
on $\partial D$.
In this section, we shall examine what happens when the variations
of $e^a$ and $\omega^a$ are restricted
at the boundary $\partial D$ of a disc.
{}From equations (2.17) and (2.18),
this will also put restrictions on the dynamical properties
of the current densities $P_a(\psi)$ and $J_a(\psi)$.
[Here we do not attach an extra index $(A)$ to
$P_a(\psi)$ and $J_a(\psi)$
since there is only one boundary $\partial D$.]
For example, in one case that we shall study
[case  c) ], the angular momentum current density
$J_a(\psi)$ is required to vanish, leading to a torsionless theory.

The requirement that the Gauss law be differentiable will again
impose certain conditions on the test
functions $\Lambda^{(0)}$ and $\Sigma^{(0)}$, but now
they will be less in number, as not all
components of $\Lambda^{(0)}$ and $\Sigma^{(0)}$ need vanish at
$\partial D$.  From eq. (2.5), $\Lambda^{(0)}_a$, for a given $a$,
need not be zero at
$\partial D$, if the value of $\omega^a|_{\partial D}$ is fixed.
Similarly, $\Sigma^{(0)}_a$ need not vanish at
$\partial D$, if the value of $e^a|_{\partial D}$ is fixed.

We shall only consider restrictions
on $e^a$ and $\omega^a$ at $\partial D$ such that the Gauss law
constraints remain first class.  We do this in order to
preserve the full set of $ISO(2,1)$ gauge symmetries, which presumably
a 2+1 dimensional topological theory of gravity should have.
(In addition, there would be technical difficulties if
some of the Gauss law constraints were second class.
This is since it would then become necessary
to compute Dirac brackets for the observables.)
Below we shall classify the possible restrictions
on $e^a$ and $\omega^a$ at $\partial D$.

First, let \underbar{$ISO(2,1)$} denote the Lie algebra
of $ISO(2,1)$, and let $t_a$ and $u_a$, $a=0,1,2$,
be a basis for the algebra.  The latter have the following Lie
brackets:
\be
[u_a,u_b]=0 \;,\quad
[u_a,t_b]=\epsilon_{abc}\;u^c \;,\quad
[t_a,t_b]=\epsilon_{abc}\;t^c \;.
\ee
Next define a Lie algebra valued test function ${\cal F}^{(0)}=$
${1 \over 2}\Bigl( \Lambda^{(0)}_a \; u^a+
\Sigma^{(0)}_a \; t^a \Bigr)$ for the Gauss law constraints.
{}From eq. (2.8), the Gauss law constraints will be first class provided:

\it 1) \rm ${\cal F}^{(0)}|_{\partial D}$
belongs to a subalgebra \underbar{$H$} of
\underbar{$ISO(2,1)$}.

\it 2) \rm $\Lambda^{(0)}_a |_{\partial D} $ and/or
$\Sigma^{(0)}_a |_{\partial D} $ vanishes for all $a$.

Ignoring boundary terms, condition \it 1) \rm is needed so that
the Poisson bracket of two Gauss law constraints equals a Gauss
law constraint.  The boundary terms in eq. (2.8) must vanish
for the same reason.
Condition \it 2) \rm insures that the first such boundary term, i.e.
${\kappa\over 4}
\int_{\partial D}\Bigl(\Sigma^{(0)a}d\Lambda'^{(0)}_a  +
\Lambda^{(0)a} d\Sigma'^{(0)}_a\Bigr)$,
vanishes.  From this condition it follows that the maximum
dimension of the subalgebra \underbar{$H$} is 3.
Condition \it 2) \rm is sufficient but
not necessary for the first boundary term
in eq. (2.8) to vanish.  The boundary term will also vanish for
${\cal F}^{(0)}|_{\partial D} ={1 \over 2}\Bigl(
\Lambda^{(0)}_- |_{\partial D}\; u_+ +
\Sigma^{(0)}_- |_{\partial D}\; t_+ \Bigr)$, when $u_+$ and $v_+$
are parallel light-like vectors, eg. $u_+=u_0+u_1$ and
$v_+=v_0+v_1$.  In this example, $u_+$ and $v_+$ are a basis for
\underbar{$H$}.  Then dim(\underbar{$H$})=2, so it is still true
that dim(\underbar{$H$})$\leq$ 3.
We shall study this exceptional case (case \it i)\rm )
at the end of this section.

We can classify all possible boundary conditions (consistent with
$g(\Lambda^{(0)},\Sigma^{(0)}) \approx 0$ being first class) by the
subalgebras \underbar{$H$} of \underbar{$ISO(2,1)$}, or
by the subgroups $H$ of $ISO(2,1)$.

For dim ($H$)=3, we have:

\bf a)  H=ISO(2) \rm.  For this case,
${\cal F}^{(0)}|_{\partial D} ={1 \over 2}\Bigl(
\Lambda^{(0)}_1 |_{\partial D}\; u^1+
\Lambda^{(0)}_2 |_{\partial D}\; u^2+
\Sigma^{(0)}_0 |_{\partial D}\; t^0 \Bigr)$, and we fix the values
of the forms $e^0$, $\omega^1$ and $\omega^2$ on $\partial D$.
In order for the last boundary term in eq. (2.8) to vanish it is
necessary that we fix the values
of $\omega^1$ and $\omega^2$ on $\partial D$ to be zero.

\bf b)  H=ISO(1,1) \rm.  For this case,
${\cal F}^{(0)}|_{\partial D} ={1 \over 2}\Bigl(
\Lambda^{(0)}_0 |_{\partial D}\; u^0+
\Lambda^{(0)}_1 |_{\partial D}\; u^1+
\Sigma^{(0)}_2 |_{\partial D} \;t^2 \Bigr)$, and
the forms $e^2$, $\omega^0$ and $\omega^1$ are held fixed
on $\partial D$.  In order for the last boundary term in
eq. (2.8) to vanish it is necessary that we fix the values
of $\omega^0$ and $\omega^1$ on $\partial D$ to be zero.

\bf c)  H=SO(2,1) \rm.  For this case,
${\cal F}^{(0)}|_{\partial D}={1 \over 2}
\Sigma^{(0)}_a |_{\partial D}\; t^a $ and
the forms $e^a$ are held fixed on $\partial D$.
In order for the last boundary term in eq. (2.8) to vanish it is
necessary that we fix the values of $e^a$ on $\partial D$ to be zero.
Then from eq. (2.18),
the angular momentum current density $J^a$ vanishes.

\bf d)\rm  ${\bf H=T}^3$ \rm (The translation group
in three dimensions).  For this case, ${\cal F}^{(0)}
|_{\partial D} ={1 \over 2} \Lambda^{(0)}_a |_{\partial D} \;u^a $ and
the forms $\omega^a$ are held fixed on $\partial D$.
In this case there are no restrictions on the values of
$\omega^a$ on $\partial D$.

When no conditions were placed on the values of $\omega^a$ and
$e^a$ on $\partial D$, $q (\Lambda,\Sigma)$, or equivalently
the $ISO(2,1)$ generators $P_a (\psi)$ and
$J_a(\psi)$, were the observables of the theory.  Let us see what
happens to these quantities for the cases  a-d).

\bf a) \rm  Here $J^0$, $P^1$ and $P^2$ have zero Poisson brackets
with $g(\Lambda^{(0)},\Sigma^{(0)})$ and hence are gauge invariant.
However they are non dynamical, because from eqs. (2.17) and (2.18),
they are weakly equal to the boundary values of
$\kappa e^0_\theta$, $-\kappa\omega^1_\theta$ and
$-\kappa\omega^2_\theta$, respectively (the latter two values
being zero).

The remaining quantities $P^0$, $J^1$ and $J^2$
do not have zero Poisson brackets with $g(\Lambda^{(0)},\Sigma^{(0)})$,
\be
 \{ g (\Lambda^{(0)},\Sigma^{(0)}), P^0(\psi)\}\;=\;
- {\kappa\over 2} \partial_\psi \Sigma^{(0)0}(r_0,\psi) \;,\qquad\qquad
\qquad
\ee
$$
 \{ g (\Lambda^{(0)},\Sigma^{(0)}), J^1(\psi)\}\;=\;
\; {\kappa\over 2} \partial_\psi \Lambda^{(0)1}(r_0,\psi)\qquad\qquad
 \qquad  $$
\be
\qquad\qquad\qquad +{1\over 2} \Sigma^{(0)}_0(r_0,\psi)\; J_2 (\psi) \;
+\;{1\over 2} \Lambda^{(0)}_2 (r_0,\psi)\; P_0 (\psi)  \;,
\ee
$$
 \{ g (\Lambda^{(0)},\Sigma^{(0)}), J^2(\psi)\}\;=\;
\; {\kappa\over 2} \partial_\psi \Lambda^{(0)2}(r_0,\psi)\qquad \qquad
\qquad$$
\be
\qquad\qquad\qquad -\;{1\over 2}\Sigma^{(0)}_0(r_0,\psi)\; J_1 (\psi)  \;
-\;{1\over 2} \Lambda^{(0)}_1 (r_0,\psi)\; P_0 (\psi) \;,
\ee
where we have written $\Lambda^{(0)a}$ and $\Sigma^{(0)a}$ as
functions of polar coordinates $r$ and $\theta$, with $r=r_0$ once
again corresponding to the boundary.
Equations (6.2-4) show that $P^0$, $J^1$ and $J^2$ are not observables.
Instead, they transform under gauge transformations as components of an
$H=ISO(2)$ connection one form.

To see this define one forms $\Omega$ and
$E^i,\; i=1,2$ on a circle parametrized by $\psi$ as follows:
\be
\Omega= -{1\over \kappa}  P^0 (\psi) d\psi  \quad
{\rm and}  \quad E^i=  {1\over\kappa}  J^i (\psi) d\psi  \;.
\ee
$\Omega$ is an $SO(2)$ spin connection, and $E^i$ are zweibein one
forms.  Under infinitesimal gauge transformations generated by
$ g (\Lambda^{(0)},\Sigma^{(0)})$,
\be
\delta \Omega = d\rho \;,
\ee
\be
\delta E^i = - \rho \epsilon^{ij} E_j
+\;\epsilon^{ij} \lambda_j \Omega + d \lambda^{i} \;,
\ee
where $\rho(\psi)= {1\over 2} \Sigma^{(0)0}(r_0,\psi)$
 parametrize infinitesimal rotations, and
$\lambda_i(\psi)= {1\over 2} \Lambda^{(0)}_i(r_0,\psi)$
 parametrize infinitesimal translations in a plane.
Eq. (6.6) leads to a nonlocal observable for this system.
It is namely the integral of $\Omega$ around the boundary.
Then from eq. (6.5), $\int_0^{2\pi}  P^0 (\psi) d\psi$, or the
``energy" associated with the boundary, is an observable.

If one enlarges the underlying manifold
(i.e., $\partial D $)
 on which $\Omega$ and $E^i$
are defined (say to $\partial D \times {\bf R}^{1}$,
where ${\bf R}^{1}$ denotes the time variable),
then it is possible to introduce an $SO(2)$ curvature two form
\be
R^{(2)}=d\Omega\;,
\ee
and a torsion two form
\be
T^{(2)i}=dE^i +\epsilon^{ij} E_j \wedge \Omega
\ee
 on the manifold.  $R^{(2)}$ is invariant under local
$ISO(2)$ transformations, while $\delta T^{(2)i}=$
$\epsilon^{ij}(\rho T^{(2)}_j -\lambda_j R^{(2)})$.
The two forms $R^{(2)}$ and $T^{(2)i}$
 have been utilized previously in a theory of two dimensional
 gravity \cite{acw}.

\bf b) \rm  This case is essentially the same as \bf a) \rm, except
instead of (6.5), we define
\be
\Omega= -{1\over\kappa}  P^2 (\psi) d\psi  \quad
{\rm and}  \quad
E^i=  {1 \over\kappa}  J^i (\psi) d\psi  \;,
\ee
where now $i=0,1$.  Further,
$\Omega$ is an $SO(1,1)$ spin connection one form, and now
$\rho$ is defined by
$\rho(\psi)=  \Sigma^{(0)2}(r_0,\psi)$ which
parametrizes infinitesimal local Lorentz transformations.
The integral of $\Omega$ is again an observable of the system,
but here it corresponds to the total momentum in the 2- direction.
$R^{(2)}=d\Omega$ is now the $SO(1,1)$ curvature two form.

\bf c) \rm  In this case, the $J^a$'s have zero Poisson brackets
with $g(\Lambda^{(0)},\Sigma^{(0)})$, and hence are gauge invariant.
However from eq. (2.8) they are weakly equal to the boundary value
of $\kappa e^a_\theta$ which is zero.

The remaining variables $P^a$ have non zero
Poisson brackets with the Gauss law constraints,
\be
 \{ g (\Lambda^{(0)},\Sigma^{(0)}), P^a(\psi)\}\;=\;
-{1\over 2} \epsilon^{abc}
\Sigma^{(0)}_b(r_0,\psi)\; P_c (\psi)  \;
\;-\; {\kappa\over 2}
\partial_\psi \Sigma^{(0)a}(r_0,\psi) \;,
\ee
and are therefore not gauge invariant.  They transform as components of
an $H=SO(2,1)$ connection one form.
The connection one form can be defined
by
\be
\Omega^a =-{1\over \kappa}  P^a (\psi) d\psi \quad.
\ee
Then under gauge transformations
\be
\delta \Omega^a =
- \epsilon^{abc} \rho_b \Omega_c
 +d\rho^{a} \;,
\ee
where $\rho^a(\psi)= {1\over 2} \Sigma^{(0)a}(r_0,\psi)$
 parametrize infinitesimal $SO(2,1)$ transformations.

If, as with case ${\bf a)}$,
 one enlarges the underlying manifold on which
$\Omega^a$ are defined to $\partial D \times {\bf R}^{1}$,
it is possible to introduce an $SO(2,1)$ curvature two form
\be
R^{(2)a}=d\Omega^a+{1\over 2}\epsilon^{abc}\Omega_b
\wedge\Omega_c\quad,
\ee
which is gauge covariant, ie.
$\delta R^{(2)a}=-\epsilon^{abc}\rho_b R^{(2)}_c$.
In this case, there is no torsion two form which we can define
on the boundary.

\bf d) \rm  Now the $P^a$'s have zero Poisson brackets
with $g(\Lambda^{(0)},\Sigma^{(0)})$, and hence are gauge invariant.
But they are weakly equal to the boundary values
of $-\kappa\omega^a_\theta$.  They are then non dynamical.

The $J^a$'s have non zero
Poisson brackets with the Gauss law constraints,
\be
 \{ g (\Lambda^{(0)},\Sigma^{(0)}), J^a(\psi)\}\;=\;
{1\over 2} \epsilon^{abc}
\Lambda^{(0)}_b(r_0,\psi)\; P_c (\psi)  \;
\;+\; {\kappa\over 2}
\partial_\psi \Lambda^{(0)a}(r_0,\psi) \;,
\ee
Now define the one forms
\be
E^a=  {1 \over\kappa}  J^a (\psi) d\psi  \quad {\rm and} \quad
\Omega^a =-{1\over \kappa}  P^a (\psi) d\psi \quad.
\ee
Then under gauge transformations
\be
\delta E^a = - \epsilon^{abc} \lambda_b \Omega_c +d\lambda^{a} \;,
\ee
where $\lambda^a(\psi)= {1\over 2} \Lambda^{(0)a}(r_0,\psi)$
 parametrize local translations.
If one enlarges the underlying manifold on which
$E^a$ and $\Omega^a$ are defined to $\partial D \times {\bf R}^{1}$,
one can define a torsion two form
\be
T^{(2)a}=dE^a+\epsilon^{abc}E_b \wedge\Omega_c   \;.
\ee
Under a local translation
$\delta T^{(2)a}=-\epsilon^{abc}\lambda_b R^{(2)}_c$,
where $R^{(2)}_a$ was defined in eq. (6.14).  Here however
it is non dynamical, since neither is $\Omega^a$.

In summary, we find that for dim $(H)=3$ the field content of the theory
consists to the set of
connection one forms associated with gauge group $H$.
These connection one forms are constructed from the (unconstrained)
$P_a$'s and $J_a$'s.  Two dimensional gravity theories can be
formulated in terms of such fields \cite{acw}.

Connection one forms for the gauge group $H$
can also be identified when dim$(H)<3$.  For these cases,
there also exist observable degrees of freedom
amongst the $P_a$'s and $J_a$'s.  The case of $H$ being the
 trivial group (containing just the identity)
 was discussed in the previous section, where
it was found that all of the $P_a$'s and $J_a$'s were observable.
When $H$ is not the trivial group, combinations of (unconstrained)
$P_a$'s and $J_a$'s can be formed which are observables, which we shall
show below.  There are
essentially five such cases to be considered.  They are:
\bf e) \rm $H=T^2$, \bf f) \rm $H=T^1$, \bf g) \rm $H=SO(1,1)$,
\bf h) \rm $H=SO(2)$ and \bf i) \rm $H=T^2_-$.
We discuss these cases in what follows.

\bf e) \rm ${\bf H=T}^2$.
In this case, ${\cal F}^{(0)}|_{\partial D} ={1 \over 2}
\Lambda^{(0)}_i |_{\partial D} \;u^i $ and we choose $i=0,1$.
For $ g (\Lambda^{(0)},\Sigma^{(0)})$ to be differentiable
the forms $\omega^i$ are held fixed on $\partial D$ and there
is no restriction on their values coming from the requirement
that $ g (\Lambda^{(0)},\Sigma^{(0)})$ is first class.  As in case
\bf d) \rm, the $P^a$'s have zero Poisson brackets
with $g(\Lambda^{(0)},\Sigma^{(0)})$, and hence are gauge invariant.
{}From eq. (2.17), they are weakly equal to the boundary values of
$-\kappa\omega^a$.  Then $P^2$ is a dynamical quantity, while $P^i
,\;i=0,1$ are constrained.

The $J^a$'s, in general, have non zero
Poisson brackets with the Gauss law constraints, as eq. (6.15)
still applies (only with $ \Lambda^{(0)}_2(r_0,\psi)=0$).
{}From $J^i,\; i=0,1$, we can construct the $T^2$
connection one forms $E^i$ according to eq. (6.16).
Under local translations,
$\delta E^i = -\epsilon^{ij} \lambda_j \Omega_2  +d\lambda^{i} $,
where $\lambda^i(\psi)= {1\over 2} \Lambda^{(0)i}(r_0,\psi)$
and $\Omega_2$ is defined in eq. (6.16).
If one again enlarges the underlying manifold on which
$E^a$ and $\Omega^a$ are defined to $\partial D \times {\bf R}^{1}$,
one can define the torsion two form eq. (6.18).
Under the action of the two dimensional translation group
$\delta T^{(2)i}=-\epsilon^{ij}\lambda_j R^{(2)}_2$,
where $R^{(2)}_a$ was defined in eq. (6.14).

In the above discussion we found that $P^2$
is an observable.  A second observable can be obtained in this
system if we set $\omega^i |_{\partial D}=0$.
This other observable is $J^2$, because
$P^i$ are weakly zero, and consequently
\be
 \{ g (\Lambda^{(0)},\Sigma^{(0)}), J^2(\psi)\}\;=\;
{1\over 2} \epsilon^{ij} \Lambda^{(0)}_i(r_0,\psi)\; P_j (\psi)  \;
\approx 0         \;.
\ee
The Poisson bracket between the
two observables $P^2$ and $J^2$ is obtained from eq. (2.24),
\be
\{P_2(\psi),J_2(\psi')\} =-\kappa \partial_\psi \delta(\psi-\psi') \;.
\ee
This relation defines an abelian Kac-Moody algebra, where the generators
are $J^1+P^1$ and $J^1-P^1$.

If we had we chosen $i$ to take the values $1,2$, instead of $0,1$,
then the above analysis would be the same, with perhaps the only
difference being that the
sign of the central term in eq. (6.20) would then be +.

\bf f) \rm ${\bf H=T}^1$.
We choose ${\cal F}^{(0)}|_{\partial D} ={1 \over 2}
\Lambda^{(0)}_2 |_{\partial D} \;u^2 $, i.e. $H$ is the group of
space translations.  Now for $ g (\Lambda^{(0)},\Sigma^{(0)})$
to be differentiable we need only require that
$\omega^2$ be held fixed on $\partial D$.  As before, there
is no restriction on its value coming from the requirement
that $ g (\Lambda^{(0)},\Sigma^{(0)})$ be first class.
Also as before, the $P^a$'s are gauge invariant.  Here, however,
$P^2$ is not dynamical, as it is weakly equal to the boundary value
of $-\kappa\omega^2_\theta$, while  $P^i,\;i=0,1$ are dynamical
quantities, and hence observables.
They have zero Poisson brackets with each other, and thus yield
a trivial algebra.

The $J^a$'s are not gauge invariant.  $J^i,\;i=0,1$, transforms
according to
\be
 \{ g (\Lambda^{(0)},\Sigma^{(0)}), J^i(\psi)\}\;=\;
-{1\over 2} \epsilon^{ij} \Lambda^{(0)}_2(r_0,\psi)\; P_j (\psi)  \;.
\ee
It follows that the product $ J^i(\psi) P_i (\psi) $ is gauge
invariant and hence observable.
{}From $J^2$, we can construct the $T^1$
connection one form $E^2$ as in eq. (6.16).  Under local translations,
$\delta E^2 = d\lambda^2 $,
where $\lambda^2(\psi)= {1\over 2} \Lambda^{(0)2}(r_0,\psi)$.
It follows that the integral of $E^2$ around the boundary is
an observable.  It corresponds to the total angular momentum
in the 2-direction.

After enlarging the underlying manifold
to $\partial D \times {\bf R}^{1}$,
one can define the torsion two form according to eq. (6.18).
$T^{(2)2}$ is invariant
under the action of the one dimensional translation group.

When ${\cal F}^{(0)}|_{\partial D} ={1 \over 2}
\Lambda^{(0)}_0 |_{\partial D} \;u^0 $, $H$ is the group of
time translations.  In that case, $P^1$, $P^2$ and $J^1P_1 +J^2P_2$
are observables, while $E^0$ is the connection one form associated
with $T^1$.

\bf g)  H=SO(1,1) \rm.
Here ${\cal F}^{(0)}|_{\partial D} ={1 \over 2}
\Sigma^{(0)}_2 |_{\partial D} \;t^2 $, and
$e^2$ is held fixed on $\partial D$.  No restriction on the value
of $e^2$ comes
 from the requirement that $ g (\Lambda^{(0)},\Sigma^{(0)})$ is
 first class.  Now, of all the $J^a$'s and $P^a$'s, only
$J^2$ is gauge invariant.  But it is
not dynamical, as it is weakly equal to the boundary value
of $\kappa e^2_\theta$.  Under a local $SO(1,1)$ gauge transformation,
both $J^i$ and $P^i, \;i=0,1$ transform as two dimensional vectors.  Thus
although they are not observables, their magnitudes are.  It
is also possible to construct additional bilinears from
$J^i$ and $P^i$, namely
$P^i(\psi)J_i(\psi) \quad{\rm and}\quad
\epsilon^{ij}P_i(\psi)J_j(\psi) \;,$ which are gauge invariant.

The connection one form
for $SO(1,1)$ is $\Omega$ as defined in eq. (6.10).
Under $SO(1,1)$ gauge transformations, $\delta \Omega = d\rho^2  \;, $
where $\rho^2(\psi)= {1\over 2} \Sigma^{(0)2}(r_0,\psi)$.
Now the integral of
$\Omega$, corresponding to the total momentum in the 2-direction,
is an observable.

After enlarging the underlying manifold to
$\partial D \times {\bf R}^{1}$, one can define
$SO(1,1)$ curvature two form according to $ R^{(2)2}=d\Omega^2$.
It is invariant under the action of the local Lorentz group.

\bf h)  H=SO(2) \rm.
Now ${\cal F}^{(0)}|_{\partial D} ={1 \over 2}
\Sigma^{(0)}_0 |_{\partial D} \;t^0 $, and
$e^0$ is held fixed on $\partial D$.
The analysis here is the same as in case \bf g) \rm, only now
$i$ takes values $1$ and $2$, and the index $2$ is replaced everywhere
by $0$.

\bf i) \rm ${\bf H=T}^2_-$.
Finally, we consider the exceptional case referred to earlier,
where the group $H$, which we denote by $T^2_-$, is generated by
the parallel light-like vectors $u_+=u_0+u_1$ and
$v_+=v_0+v_1$, and is thus abelian.
We write, ${\cal F}^{(0)}|_{\partial D} ={1 \over 2}\Bigl(
\Lambda^{(0)}_- |_{\partial D}\; u_+ +
\Sigma^{(0)}_- |_{\partial D}\; t_+ \Bigr)$, with $e_+=e_0+e_1$
and $\omega_+=\omega_0+\omega_1$ held fixed on $\partial D$.
$ g (\Lambda^{(0)},\Sigma^{(0)})$ is then differentiable, and
no restrictions on the values of $e_+$ and $\omega_+$ come
 from the requirement that it be first class.
$J^-=J^0-J^1$ and $P^-=P^0-P^1$
are invariant under gauge transformations,
but they are weakly equal to
$\kappa  e^-_\theta=\kappa (e^0_\theta-  e^1_\theta)$
and $-\kappa \omega^-_\theta =-\kappa(\omega^0_\theta
-\omega^1_\theta)$, respectively, and hence are
not dynamical.  The remaining $P^a$'s and $J^a$'s are, in general,
 not gauge invariant.  However,
 $P^2$ will be gauge invariant if $\omega_+$
vanishes on the boundary, and, in addition, $J^2$ will be gauge
invariant if $e_+$ vanishes there as well.  In the latter case, we
recover the abelian Kac-Moody algebra eq. (6.20) for the two observables.

The connection one forms for  $T^2_-$ are
$\Omega^+=\Omega^0 +\Omega^1$ and $E^+=E^0+E^1$.
Under gauge transformations,
$\delta \Omega^+ =2\rho_-\Omega_2 +d\rho_-   $ and
$\delta E^+ =2\lambda_-\Omega_2 +2\rho_-E_2 + d\lambda_-  \;, $
where $\rho_-(\psi)= {1\over 2} \Sigma^{(0)}_-(r_0,\psi)$ and
$\lambda_-(\psi)= {1\over 2} \Lambda^{(0)}_-(r_0,\psi)$ parametrize the
$T^2_-$ transformations.  The corresponding curvature and torsion
two forms are $R^{(2)+}$ and $T^{(2)+}$.  Under gauge transformations,
$\delta R^{(2)+} =2\rho_-R^{(2)}_2 $ and
$\delta T^{(2)+} =2\lambda_-T^{(2)}_2 $.  \\

\sxn{CONCLUDING REMARKS}

We believe that the formalism developed here for treating $2+1$ gravity
on manifolds with boundaries has the virtue of simplicity.
Now we outline some possible implications and
extensions of our work.

\bf The spin-statistics theorem.  \rm
In Section 4, point particles were obtained by taking the zero size
limit of holes on the disc.  As a result, an entire $ISO(2,1)$ Kac-Moody
algebra was associated with each point particle.  It would be of
interest to study the question of the spin-statistics theorem from this
point of view.  The spin-statistics theorem in $2+1$ gravity has been
examined by a number of authors, from a number of points of view
\cite{car,ks,cap}.
It was studied by two of us in ref.
\cite{ks}, but there, point particles were only labeled by their charges,
the particle momenta $p^a$ and angular momenta
$j^a$, and the very important question of self-interactions
was neglected.
With the formalism developed here, however, self-interactions of the
particle are included in a natural way.

In ref. \cite{cs2}, the spin-statistics theorem was proved for particles
in general Chern-Simons theories, where the particles
are gotten by acting on the vacuum
with the vertex operator.  (The particles are therefore associated with
the state $V|0>$.)
It appears that this result can be lifted
straightforwardly to apply to Chern-Simons gravity.
In ref. \cite{cs2}, the Virasoro charge $L_0$ played
the role of the rotation generator, as it was responsible for global
deformations of the disc.  The eigenvalue of $L_0$ was shown
to be identical to the phase associated with a two-particle exchange,
hence proving the spin-statistics connection.

But in Chern-Simons gravity, there is a another kind of rotation
generator, namely, the particle angular momentum operator
$\int_0^{2\pi}d\psi\; {\bf J}^{0}(\psi)$, which generates rotations
in the internal vector space.  (Here we ignore the particle index $(A)$.)
A question then arises as to whether the spin-statistics theorem
should be formulated in terms of $L_0$ or the particle angular momentum.
Happily, both of these operators
are equivalent when acting on the state $V|0>$, and when $n=1$.
To see this, let us promote the expression
(2.31) for the Virasoro generators to an operator equation,
\be
L(v)=
-{1\over\kappa}\int_0^{2\pi}d\psi\;\epsilon(\psi)
{\bf J}_a(\psi){\bf P}^a(\psi) \;.
\ee
With the ordering as shown in
eq. (7.1), $L(v)$ has a well defined action on the states \cite{sss}.
In particular, when acting on the state $V|0>$, we have
\be
L(v)\;V|0>=
n\int_0^{2\pi}d\psi\;\epsilon(\psi) {\bf J}^0(\psi)\;V|0> \;,
\ee
where we have used eq. (5.11).
The Virasoro charge $L_0$ is obtained by setting $\epsilon(\psi)$
in $L(v)$ equal to a constant.
The correct normalization is $\epsilon(\psi)=1$. So for the case
$n=1$, $L_0$ is equal to
the angular momentum operator when acting on
$V|0>$.  (In Sec. 5, we assumed the trivial representation for the little
group of $|0>$, and
hence also for the little group of $V|0>$.  Then both $L_0$ and
$\int_0^{2\pi}d\psi\; {\bf J}^{0}(\psi)$ are zero on $V|0>$.
The assumption of triviality may
however be dropped for the purpose of the discussion here.)
For the case $n>1$, there is an $n$ to $1$ map from the set of rotations
$\{e^{i\theta L_0}\}$ to $\{e^{i\theta\int_0^{2\pi}d\psi\;
{\bf J}^{0}(\psi)}\}$.

\bf Cylindrical space-time.  \rm  Although much is known about
conical space-time, the space-time created by particles
in $2+1$ dimensions
with mass $m<2\pi\kappa$, not much seems to be known
about cylindrical space-times \cite{djt}.
Once again, it is the latter that are created
by our vertex operator and they
are associated with particles whose masses
are given in eq. (5.13).  The scattering of such particles may be
of interest, along with their possible relevance for cosmic strings.

\bf Alternative boundary conditions. \rm   In Sec. 6, we classified
all possible boundary conditions for $2+1$ gravity on a disc.
A rich structure was found, which in several instances, yielded
the field content of $1+1$ gravity.  Thus it seemed possible to
generate the lower dimensional gravity theory.
However, the dynamics of the
lower dimensional theory seemed to be lacking.
The questions then remain as to how to introduce dynamics in
a natural manner, and further how to quantize the resulting
theory.

If there is no unique way to treat the boundary of a
disc, then there is also no unique way to treat the boundaries of holes
on the disc.  Since
point particles, here, result from shrinking holes to
points, there can be alternative descriptions of point
sources as well.  The classification of all such point sources which one
can have in $2+1$ gravity should be of considerable interest.

\newpage

{\bf Acknowledgements}

We are indebted to A. P. Balachandran for discussions during all phases
of this work.  We also wish to thank the
 group in Naples and Giuseppe Marmo, in particular, for their
 hospitality while this work was initiated.
We have been supported during the course of this work as follows:~1)
G.~B. and~K.~S.~G.~ by the Department of Energy, USA,
under contract number DE-FG2-85ER-40231, and
A.~S.~by the Department of Energy, USA under contract
number DE-FG05-84ER40141;
2) A.~S.~ by INFN, Italy [at Dipartimento di
Scienze Fisiche, Universit{\`a} di Napoli]; 3) G.~B.~ by the Dipartimento
di Scienze Fisiche, Universit{\`a} di Napoli.


\newpage

\begin{thebibliography}{99}

\bibitem{wit} E. Witten, Nucl. Phys. {\bf B311} (1988) 46;
 {\bf B323} (1989) 113.  See also, A. Achucarro and P. K. Townsend,
Phys. Lett. {\bf B180} (1986) 89.

\bibitem{wit2} E. Witten, Commun. Math. Phys. {\bf 121} (1989) 351.

\bibitem{bos} G. V. Dunne, R. Jackiw and C. A. Trugenberger,
Ann. Phys. {\bf 194}, 197 (1989)
; K. Gawedzki, IHES/P/89/06; M. Bos
and V. P. Nair, Phys. Lett. {\bf B223}, 61 (1989);
Int. J. Mod. Phys. {\bf A5}, 959 (1990);
T. R. Ramadas, Comm. Math. Phys. {\bf128},
421 (1990);
A. P. Polychronakos, Ann. Phys. {\bf 203}, 231 (1990);
D. Boyanovsky, E. T. Newman and C. Rovelli, University of
Pittsburgh preprint PITT-91-14 (1991).

\bibitem{bal} A. P. Balachandran, M. Bourdeau
and S. Jo,  Mod. Phys. Lett. {\bf A4}, 1923 (1989); Int. J. Mod. Phys.
{\bf A5}, 2423 (1990) [Erratum : ibid. 3461, (1990)].

\bibitem{ks} K. S. Gupta and A. Stern, Phys. Rev. {\bf D44} (1991) 2432.

\bibitem{moore} G. Moore and N. Seiberg,  Phys. Lett. {\bf B220},
422 (1989);  S. Elitzur, G. Moore, A. Schwimmer and N. Seiberg,
Nucl. Phys. {\bf B326},
108 (1989); A. Schwimmer in ``{\it Recent Developments in Conformal Field
Theories}'', edited by S. Randjbar-Daemi, E. Sezgin and J. B. Zuber
 (World Scientific, Singapore, 1989).

\bibitem{lee} T. R. Ramadas, I. M. Singer and J. Weitsman,
Comm. Math. Phys. {\bf 126}, 406 (1989);
L.~Smolin, Mod. Phys. Lett. {\bf A4}, 1091 (1989).

\bibitem{guad} E. Guadagnini, M. Martellini and M. Mintchev,
Nucl. Phys. {\bf B330},
575 (1990); Phys. Lett. {\bf B235}, 275 (1990);
Nucl. Phys. {\bf B336}, 581 (1990).

\bibitem{cs1} A. P. Balachandran, G. Bimonte, K. S. Gupta and
A. Stern,``Conformal Edge Currents in Chern-Simons Theories,''
talk presented
by K. S. Gupta at the Montreal-Rochester-Syracuse-Toronto meeting,
Rochester (1991), Syracuse University preprint SU-4228-
477, INFN-NA-IV-91/22, UAHEP 915 (1991) and
Proceedings of the Thirteenth Annual
Montreal-Rochester-Syracuse-Toronto meeting (1991) page 25;
A. P. Balachandran, G. Bimonte, K. S. Gupta and A. Stern,
University of Alabama, Syracuse University and
Universit{\`a} di Napoli preprint
UAHEP 9113, SU-4228-487,
INFN-NA-IV-91/12, October (1991), and Int. J. of Mod. Phys. {\bf A}
(in press).

\bibitem{cs2} A. P. Balachandran, G. Bimonte, K. S. Gupta and
A. Stern,``The Chern-Simons Source as a Conformal Family and its
Vertex Operators'',
Syracuse University, Universit{\`a} di Napoli and
University of Alabama preprint
SU-4228-491, INFN-NA-IV-91/13, UAHEP 9114, January (1992),
and Int. J. of Mod. Phys. {\bf A} (in press).


\bibitem{zhang} S. C. Zhang, T. H. Hansson, S. Kivelson , Phys.
Rev. Lett. {\bf 62}, 82 (1989);
F. Wilczek
``{\it Fractional Statistics and Anyon Superconductivity }''
(World Scientific, Singapore, 1990) and articles therein;
G. Moore and N. Read, Yale preprint YC TP-P6-90 (1990);
J. Frohlich
and T. Kerler, Nucl. Phys. {\bf B354}, 369 (1991);
G. Cristofano, G. Maiella, R. Musto and F. Nicodemi,
Mod. Phys. Lett. {\bf A6},
{}~2985 (1991); M. Stone and H. W. Wyld, University of Illinois, Urbana
preprint
ILL-TH-91-21 (1991)
and references in these papers.

\bibitem{blok} B. Blok and X. G. Wen, Institute for Advanced Study,
Princeton  preprint
IASSNS-HEP-90/23 (1990); X.G.Wen, ibid, IASSNS-HEP-91/20 (1991);
J. Frohlich and A. Zee,
Institute for Theoretical Physics, Santa Barbara preprint
ITP-91-31 (1991) and references in these papers;
A. P. Balachandran and A. M. Srivastava, Minnesota and Syracuse
preprint, TPI-MINN-91/38-T, SU-4228-492.

\bibitem{go} For a review, see P. Goddard and D. Olive,
Int.~J.~Mod.~Phys.~{\bf A1} (1986) 303.

\bibitem{sss} A. Stern, ``Strings on the $ISO(2,1)$ Manifold",
in {\it Superstrings and Particle Theory},
eds. L. Clavelli and B. Harms,
World Scientific, Singapore, 1989;
 P. Salomonson, B.-S. Skagerstam and A. Stern, Nucl. Phys.
{\bf B347} (1990) 769;
A. Stern, Int. J. of Mod. Phys. {\bf A6} (1991) 5215.


\bibitem{wig} E. Wigner, Ann. Math. {\bf 40} (1939) 149;
Also see A. P. Balachandran and C. G. Trahern, ``{\it Lectures on Group
Theory for Physicists}", Part IV, (Bibliopolis, edizioni di filosofia
e scienze, Napoli, 1984).

\bibitem{djt} S. Deser, R. Jackiw and G. 't Hooft, Ann. of Phys.
 {\bf 152} (1984) 220.
\bibitem{sta}
A. Starnszkiewicz, Acta. Phys. Pol. {\bf 24}, (1963), 734;
    J.R. Gott and M. Alpert, Gen. Relativ. {\bf 16} (1984) 751;
    P.O. Mazur, Phys. Rev. Lett. {\bf 57} (1986) 929
    R. Jackson, MIT preprint CTP 1824 (1989):
    S. Deser, Phys. Rev. Lett. {\bf 64} (1990) 611;
    P. de Sousa Gerbert and R. Jackiw, Commun. Math. Phys. {\bf 124}
    (1989) 226;
    D. Harari and AP. Polychronakos, Phys. Rev. {\bf D38} (1988) 3320.

 \bibitem{gri}
    G. Grignani and G. Nardelli, MIT preprints CTP 1946,1953,2001,2017;
    A. Capelli, M. Ciafaloni and P. Valtancoli, CERN preprint
    CERN-TH-6248/91.
    C. Vaz and L. Witten, Cincinatti preprints UCTP-103/91,
    UCEH-101/92.

\bibitem{ss} B.-S. Skagerstam and A. Stern, Int. J. of Mod. Phys.
{\bf A5} (1990)1575;   P. de Sousa Gerbert, Nucl. Phys. {\bf B346}(1990)
440.  Also see A.P. Balachandran, G. Marmo, B.-S. Skagerstam and A.
    Stern, Phys. Lett. {\bf 898} (1980) 199; ``{\it Gauge Symmetries and
    Fibre Bundles-Applications to Particle Dynamics}", Lecture
    Notes in Physics, {\bf 188} (Springer-Verlag, 1983).

\bibitem{car} S. Carlip, Nucl. Phys. {\bf B324} (1989) 106;
 Princeton preprint IASSNS-HEP-89151.

\bibitem{cap}
    A. Capelli, M. Ciafaloni and P. Valtancoli, CERN preprint
    CERN-TH-6093/91.

\bibitem{acw}
A. H. Chamseddine and D. Wyler, Phys. Lett. {\bf B228} (1989) 75;
M. A. Awada and A. H. Chamseddine, Phys Lett. {\bf B233} (1989) 79;
A. H. Chamseddine and D. Wyler, Nucl. Phys. {\bf B340} (1989) 595;
D. Montano and J. Sonneschein, Nucl. Phys. {\bf B324} (1989) 348;
K. Isler and C. Trugenberger, Phys. Rev. Lett. {\bf 63} (1989) 834;
M. Blau and G. Thompson, Ann. Phys. {\bf 205} (1991) 130;
L. F. Cugliandolo, F. A. Schaposnik and V. Vucetich, ICTP preprint
 IC/91/184.
\end{thebibliography}
\end{document}